\documentclass[fleqn,usenatbib]{mnras}
\usepackage{newtxtext,newtxmath}
\usepackage{float}
\usepackage{color}
\usepackage{xcolor}
\usepackage{booktabs}
\usepackage{multirow}
\usepackage{natbib}
\usepackage{comment}
\usepackage[normalem]{ulem}
\usepackage[T1]{fontenc}
\usepackage{ae,aecompl}
\usepackage{gensymb}
\usepackage{soul}

\usepackage{caption,subcaption}

\usepackage{graphicx}	
\usepackage{amsmath}	
\usepackage[export]{adjustbox}[2011/08/13]

\usepackage{mathrsfs}

\newcommand{\Ms}{M$_\odot $}

\newcommand{\vir}[1]{``#1''}

\newcommand{\mob}{\texttt{MOBSE}}

\definecolor{seagreen}{rgb}{0.090, 0.725, 0.185}

\begin{document}
\title[BBHs in YSCs]{Dynamics of binary black holes in low-mass young star clusters }


\author[Rastello et al.]{Sara Rastello$^{1,2}$\thanks{E-mail: sara.rastello@unipd.it},  Michela Mapelli$^{1,2,3}$\thanks{E-mail: michela.mapelli@unipd.it}, Ugo N. Di Carlo$^{1,2,3}$,
Giuliano Iorio$^{1,2,3}$,  \newauthor  Alessandro Ballone$^{1,2,3}$,     Nicola Giacobbo$^{1,4}$,  Filippo Santoliquido$^{1,2}$, Stefano Torniamenti$^{1,2,3}$
\\
$^{1}$Physics and Astronomy Department Galileo Galilei, University of Padova, Vicolo dell'Osservatorio 3, I-35122 Padova, Italy\\
$^{2}$INFN - Padova, Via Marzolo 8, I--35131 Padova, Italy\\
$^{3}$INAF - Osservatorio Astronomico di Padova, Vicolo dell'Osservatorio 5, I-35122 Padova, Italy\\
$^{4}$School of Physics and Astronomy, Institute for Gravitational Wave Astronomy, University of Birmingham, Birmingham, B15 2TT, UK}

\date{Accepted XXX. Received YYY; in original form ZZZ}

\pubyear{2020}

\maketitle

\begin{abstract} 

Young star clusters are dynamically active stellar systems and are a common birthplace for massive stars. Low-mass star clusters ($\sim{}300-10^3$ M$_\odot$) are more numerous than massive systems and are characterized by a two-body relaxation time scale of a few Myr: the most massive stars sink to the cluster core and dynamically interact with each other even before they give birth to compact objects. Here, we explore the properties of black holes (BHs) and binary black
holes (BBHs) formed in low-mass young star clusters, by means of a suite of $10^5$ direct $N$-body
simulations with a high original binary fraction (100\% for stars with mass $>5$ M$_\odot$). Most BHs are ejected in the first $\sim{}20$ Myr by dynamical interactions. Dynamical exchanges are the main formation channel of BBHs, accounting for $\sim{}40-80$\% of all the systems.  Most BBH mergers in low-mass young star clusters involve primary BHs with mass $<40$ M$_\odot$ and low mass ratios are extremely more common than in the field. Comparing our data with those of more massive star clusters ($10^3-3\times{}10^4$ M$_\odot$), we find a strong dependence of the percentage of exchanged BBHs on the mass of the host star cluster. In contrast, our results show just a mild correlation between the mass of the host star cluster and the efficiency of BBH mergers.

\end{abstract}



\section{Introduction} 
\label{intro}

The number of gravitational-wave (GW) events observed by the LIGO--Virgo collaboration (LVC, \citealt{LIGOdetector,Virgodetector}) has already grown to several dozens. In particular, the second GW transient catalogue (GWTC-2, \citealt{abbottO3a,abbottO3apopandrate,abbottO3aGR}) consists of 50 candidate compact binary mergers: 3 from the first observing run (O1, \citealt{abbottGW150914,abbottGW151226,abbottO1}), 8 from the second (O2, \citealt{abbottO2,abbottO2popandrate}) and 39 from the first half of the third observing run (O3a).

Most events in GWTC-2 are associated with binary black hole (BBH) mergers with component masses
ranging between $\approx{}5$ and $\approx90$ \Ms.
GWTC-2 also includes the most massive  binary merger observed to date, GW190521 \citep{abbottGW190521,abbottGW190521astro}. In this case,
the merger of two massive black holes (BHs, $\approx85$ \Ms{} and $\approx66$ \Ms{}) resulted in the
formation of a BH remnant of $\approx142$ \Ms{} providing the first clear
detection of an intermediate-mass BH
 in the $100-1000$ \Ms{} range \citep{abbott2020b,abbott2020e,nitz2021,fishbach2020}. 
 In addition, O3a witnessed the first
observation of a BBH with asymmetric masses \citep[GW190412,][]{abbottGW190412}, and the second binary neutron star merger \citep[GW190425,][]{abbottgw190425}. Finally, GW190814 might be the first BH--neutron star binary merger, with masses $\approx{}23$ and $\approx{}2.6$ \Ms{}: its secondary component is either the lighter BH or the most massive neutron star ever observed.

This wealth of GW observations opens a new window on the study of BBHs.
 As pointed
out by various authors \citep{fishbach2017, zevin2017,stevenson2017,farr2017,vitale2017,gerosa2017,
gerosa2018, bouffanais2019,bouffanais2020,bouffanais2021,wong2019,wong2020,zevin2020,doctor2020,kimball2020,ng2020}, a few hundreds of GW detections might be
sufficient to disentangle the main formation pathways of BBHs, such as isolated binary evolution and dynamical assembly. 
On the one hand, the evolution of massive isolated stellar binaries can lead to the formation of tight BBHs either via common envelope (e.g.
\citealt{tutukov1973, bethe1998,portegieszwart1998,belczynski2002, belczynski2008,belckzynski2010,dominik2012, dominik2013,
mennekens2014, belczynski2016, loeb16,belczynski2016,demink,marchant,
mapelli2018,mapelli2019, giacobbo2018b,
kruckow2018,spera2018,tang2019,belczynski2020,garcia2021}), via chemically
homogeneous evolution \citep{demink2016,mandel2016,marchant2016,dubuisson2020}, or via stable mass transfer \citep[e.g.,][]{giacobbo2018,neijssel2019,bavera2020}. On the other hand, dynamical interactions can trigger the formation of BBHs in dense stellar systems such as
globular clusters 
\citep[GCs, e.g.,][]{downing2010,BenacquistaDowning2013,rodriguez15,rodriguez2016,antonini2016,
askar2017, fujii2017,askar2018,fragione2018,rodriguez2019}, nuclear
star clusters
\citep[e.g.,][]{oleary2009,millerlauburg2009,mckernan2012,megan17,mckernan2018,vanlandingham2016,
stone2017,hoang2018,arcagualandris2018,antonini2018,arcasedda2019,arcasedda2020,mapelli2021} and young massive star clusters
\citep[YSCs, e.g.,][]{portegieszwart2002,sambaran10,mapelli2013,ziosi2014,
goswami2014, mapelli2016, banerjee2017, banerjee2018,rastello2018,perna2019,
dicarlo2019a,kumamoto2019,kumamoto2020,rastello2020,banerjee2020}. Dynamical mergers can also be the result of (triple-multiple) hierarchical systems in the field
\citep[e.g.,][]{antonini2017,silsbee2017,frlgk18,fragione19a,fragione19b,fragione2020} or gas-assisted hierarchical assembly in AGN discs \citep[e.g.,][]{mckernan2012,mckernan2018,bartos2017,yang2019,tagawa2020}.

Isolated binary evolution predicts the formation of BBHs with 
preferentially equal-mass systems, mostly aligned spins and zero
eccentricity in the LVC band (\citealt{mandel2016,gerosa2018}, but see \citealt{stegmann2020} for a possible spin flip mechanism). 
In contrast, the dynamical formation in star
clusters might lead to even larger primary masses (e.g.
\citealt{mckernan2012,mapelli2016,antonini2016,gerosa2017,stone2017,mckernan2018,dicarlo2020,rodriguez2019,yang2019,arcasedda2019,arcasedda2020}),
mass ratios ranging from $q\sim{}0.1$ to $q\sim{}1$ (e.g.
\citealt{dicarlo2019a}), isotropic spin distribution \citep[e.g.,][]{rodriguez2016b}, and, in some rare but not
negligible cases, non-zero eccentricity in the LVC band (e.g.,
\citealt{samsing2018,samsing2018b,samsing2018c,rodriguez2018,zevin2019}).

Here, we study the formation and evolution of BBHs in extremely low-mass young star clusters
(YSCs, $300\,<\,m_{\rm{SC}}\,<\,1000$ \Ms{}).
Such small star clusters are significantly more numerous  
than the massive ones,  given the slope of the star cluster mass function (${\rm d}N/{\rm d}m_{\rm SC}\propto{}m_{\rm SC}^{-2}$, \citealt{portegieszwart2010,fuji2016}), but have frequently been  overlooked in the study of binary compact objects.

In star clusters, dynamical evolution is driven by two-body 
relaxation, which proceeds over a timescale \citep{bt}
\begin{eqnarray}\label{eq:trlx}
\centering
t_{\rm rlx}\sim{}6\,{}{\rm Myr}\,{}\left(\frac{m_{\rm SC}}{1000\,{}{\rm M}_\odot}\right)^{1/2}\,{}\left(\frac{r_{\rm h}}{1\,{}{\rm pc}}\right)^{3/2}\,{}\left(\frac{\langle{}m\rangle{}}{1\,{}{\rm M}_\odot}\right)^{-1},
\end{eqnarray}
where $r_{\rm h}$ is the cluster half mass radius and $\langle{}m\rangle{}$ is the average mass of a star in the cluster.
For low-mass young star clusters, $t_{\rm rlx}$ ranges between a few Myr and $\approx 10$ Myr, while
for massive  star clusters, 
such as globular and nuclear star clusters, $t_{\rm rlx}$ is of the order of several hundred Myr. 

Since the dynamical friction time for a star with mass $m$ scales as $t_{\rm DF}\propto{}t_{\rm rlx}\,{}\langle{}m\rangle{}/m$ \citep{cha43i}, massive stars usually die before they can sink to the centre of a globular/nuclear star cluster, while they can reach the core of a low-mass YSC before they collapse to BH \citep{portegieszwart2000}. Hence, mass segregation is extremely efficient in low-mass YSCs, leading to frequent dynamical interactions between massive stars in the first Myrs 
\citep[e.g.,][]{portegieszwart2002,fujii2014}: the stellar progenitors of BHs and neutron stars can even collide with each other in the core of a low-mass YSC, before collapsing to compact objects \citep[e.g.,][]{kumamoto2019,dicarlo2019a}.

Given the low velocity dispersion of YSCs (just a few km s$^{-1}$, \citealt{spitzer1987,portegieszwart2010,rastello2020apj}), most binary stars are hard (i.e., they have binding energy larger than the average kinetic energy of a cluster star, \citealt{heggie1975}) in such environments. This implies that almost all the original binary stars survive in a small YSC, while in nuclear star clusters most binaries are soft and get disrupted by encounters. Hence, dynamical exchanges involving original binaries are favoured in low-mass YSCs with respect to more massive clusters. For these reasons, the dynamical processes affecting their stellar progenitors (e.g., mass segregation, stellar collisions and dynamical exchanges) play a more important role for the formation of BBHs in low-mass YSCs than in more massive clusters. 

Another peculiarity of these systems is that each cluster hosts a few BH and neutron star progenitors ($\sim{}1-6$ for star clusters with mass $\sim{300-1000}$ \Ms). This favours the pair up of BH--neutron star systems with respect to more massive clusters \citep{rastello2020}.  Low-mass YSCs are extremely  clumpy
and asymmetric stellar systems \citep{ballone2020}, almost impossible to model with Monte Carlo codes, which are based on the assumption of spherical symmetry  \citep{henon}.  Moreover, stellar and binary evolution are a key
ingredient in the life of YSCs. Since they evolve on timescales 
comparable to
 the lifetime of their massive stars (eq.~\ref{eq:trlx}), mass loss by supernova (SN) explosions and by stellar winds 
 contribute significantly to their dynamical evolution 
\citep{mapellibressan2013,trani2014}. Hence, in order to model low-mass YSCs, we have to run expensive direct $N$-body simulations coupled with binary population-synthesis codes.

Here, we make use of the 
the dynamical code \texttt{NBODY6++GPU} \citep{wang2015,wang2016}, coupled with \texttt{MOBSE} \citep{mapelli2017,giacobbo2018}.

\section{Methods}
\label{met_mod}

\subsection{Numerical Codes}

We performed a suite of  $N$-body simulations
using the direct-summation $N$-body code \texttt{NBODY6++GPU} \citep{wang2015}.
In particular, we adopted the same version of the code 
 as described in \cite{dicarlo2019a}.

\texttt{NBODY6++GPU} is the GPU parallel version of \textsc{nbody6}
\citep{aarseth2003}. It implements a 4th-order Hermite integrator, individual
block time–steps \citep{makino1992} and Kustaanheimo-Stiefel 
regularization of close encounters and few-body systems
\citep{stiefel1965,kschain}. Post-Newtonian terms are not included 
in the version of the code we used.
A neighbour scheme \citep{nitadori2012} is used to compute the force
contributions at short time steps (\textit{irregular} force/timesteps)
while for longer time steps (\textit{regular} force/timesteps) all the
members of the system are included when evaluating the force. 
The regular forces are computed on GPUs using
the CUDA architectures while irregular forces
are evaluated using CPUs.

In our simulations, \texttt{NBODY6++GPU} is interfaced with the binary population synthesis code \mob{} \citep{mapelli2017,giacobbo2018,giacobbo2018b},
which is an upgrade of \texttt{BSE} \citep{hurley2000,hurley2002}, including up-to-date
prescriptions for stellar winds, electron capture, pair instability, pulsational pair instability and 
 core-collapse SNe. Mass loss by
stellar winds is described as $\dot{M}\propto{}Z^\beta{}$ for all hot massive
stars. The index $\beta{}$  is a function of the electron-scattering Eddington ratio, accounting for
the increase of  the mass-loss rate when a star is close to its Eddington limit \citep{graefener2008,chen2015}.

We adopt the rapid core-collapse SN model \citep{fryer2012}, which prevents the formation of
compact objects with mass in the range $2-5$ M$_\odot$. 
Electron capture SNe are modelled as described in \cite{giacobbo2018c}, while (pulsational) pair instability SNe follow the formalism presented in \cite{spera2017} and
\cite{mapelli2020}. This setup produces a mass gap in the BH mass
spectrum between $m_{\rm BH}\sim{}65$~M$_\odot$ and $m_{\rm BH}\sim{}120$ M$_\odot$.

 Binary evolution processes, such as tides, mass transfer, common envelope and
GW orbital decay, are implemented as in \cite{hurley2002}. For common envelope, we adopt the energy formalism \citep{webbink1984}, assuming $\alpha{}=5$, while the concentration parameter $\lambda{}$ is calculated self-consistently as
described in \cite{claeys2014}.
 
Natal kicks are randomly drawn from a
 Maxwellian velocity distribution with 
a one-dimensional root mean square velocity $\sigma{}=15$~km~s$^{-1}$ \citep{giacobbo2018b}. For BHs, we also modulate the kick magnitude by the amount of fallback, as described in \cite{fryer2012}.
When two stars merge, the amount of mass
loss is decided by the population synthesis code \textsc{mobse}, 
which adopts the same prescriptions as \textsc{bse}. If a star merges
with a neutron star or a BH, the mass of the star is immediately lost by the merger product and
there is no mass accretion on the compact object \citep{giacobbo2018}.

\subsection{Initial Conditions}\label{sec:init_cond}

We explore three different metallicities: $Z = 0.02$, $0.002$ and
$0.0002$. We ran $33334$ direct $N-$body simulations per each metallicity for a total
of $100002$ simulations. The YSCs studied in this work have masses 
sampled in the range $300\leq{} m_{\rm SC}/{\rm M}_\odot <1000 $
from a power-law distribution $dN/dm_{\rm SC}\propto m_{\rm SC}^{-2}$, reminiscent of the
distribution of YSCs in Milky-Way like galaxies \citep{lada2003}.

The initial star cluster half mass
radius $r_{\rm h}$ is chosen according to \cite{markskroupa12}:

\begin{eqnarray}
\centering
r_{\rm
h}=0.10^{+0.07}_{-0.04}\,{}{\rm pc}\,{} \left(\frac{m_{\mathrm{SC}}}{{\rm
M}_{\odot}}\right)^{0.13\pm 0.04}.
\end{eqnarray}

Our YSCs are initially set with $q_{\mathrm{vir}} = T/|V| = 0.5$ (where $T$ and $V$ are the total kinetic and potential energy of the YSC, respectively) 
and are modelled with  fractal substructures, to account for the  initial clumpiness and asymmetry of YSCs \citep{goodwin2004}.  We choose a fractal dimension $D=1.6$, 
 as suggested by observations
\citep{sanchez2009,obs11,kuhn19} and hydrodynamical simulations
\citep{ballone2020}. We generated these initial conditions using the code
\textsc{McLuster} \citep{kuepper2011}.

Stellar masses are extracted from a Kroupa \citep{kroupa2001} initial mass
function (IMF) in the range $0.1 \le{} m \le{} 150$ \Ms. We draw the orbital parameters of original binaries following
the distributions by \cite{sana2012}. In particular, orbital periods $P$ follow the distribution
$\mathcal{P}(\Pi)\propto{}\Pi^{-0.55}$, where $\Pi\equiv{}\log_{10}(P/\mathrm{days})$
and $0.15\leq{}\Pi\leq{}6.7$, while binary eccentricities $e$
are randomly drawn from a distribution $\mathcal{P}(e)\propto{}e^{-0.42}$ with
$0\leq{}e<1$.

\begin{figure*}
\centering
\includegraphics[width=0.99\textwidth]{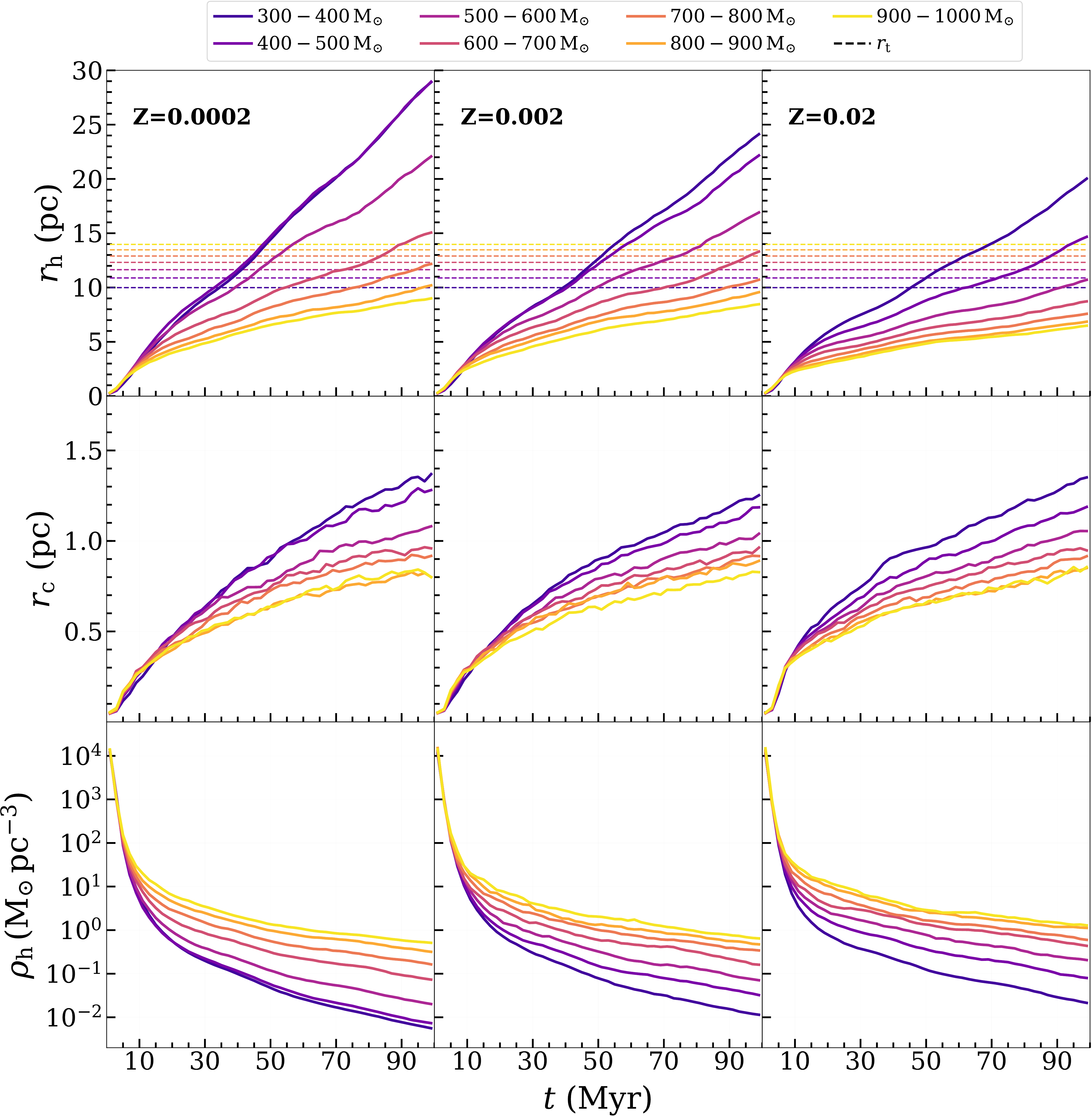}
\caption{Half-mass radius $r_{\rm h}$ (solid lines in the upper panels), tidal radius $r_{\rm t}$ (dashed lines in the upper panels), core radius $r_{\rm c}$ (solid lines in the middle panel) and density at the half-mass radius $\rho_{\rm h}$ (solid lines in the lower panels) as a function of time. 
From left to right: metallicity
$Z=0.0002$, $0.002$ and $0.02$, respectively.
  The various colours refer to different mass bins, corresponding to $m_{\rm SC}\in{}[300,400]$, $[400,500]$, $[500,600]$, $[600,700]$, $[700,800]$, $[800,900]$ and
 $[900,1000]$~M$_\odot$, respectively. 
Each line shows the median value over all the simulated YSCs, per each mass bin.
}
\label{fig:9plot}
\end{figure*}
%

We assume that initially each YSC hosts a fraction $f_{\rm bin}=0.4$ of original binaries\footnote{Here and in the following, \emph{original binaries} are binary systems already present in the initial conditions.}. 
\textsc{McLuster} \citep{kuepper2011} assigns the companion stars based on
mass: stars are randomly paired by using a distribution
$\mathcal{P}(q)\propto{}q^{-0.1}$, where $q=m_2/m_1$ is the ratio of the mass between
the secondary and the primary star according to \cite{sana2012}.
  Hence, all the stars with mass
$\,m\,\ge{}5$~\Ms{} are  members of binary systems, while stars with mass
$m\,<\,5$~\Ms{} are randomly paired until the imposed total binary fraction
$f_{\mathrm{bin}}=0.4$ is reached. The result of this method is that the most massive
stars (down to $5$ \Ms) are all binary members, while
the fraction of binaries falls to lower values for lighter stars, in agreement with \cite{moe2017}.
We embed the simulated YSCs in a solar neighbourhood-like
static external tidal field and we put them on a circular orbit around the centre
of the Milky Way at a distance $8\,\mathrm{kpc}$ \citep{wang2016}.
The BBHs that escape
 from the cluster\footnote{We define escapers as
those stars and binaries that reach a distance from the centre of the YSC
larger than twice the tidal radius, $r_{\rm{t}}$, of the cluster \citep{aarsethnb7,wang2015}.} evolve only due to the emission of gravitational radiation. 
We estimate the coalescence timescale of these binaries with the formalism
described in \cite{peters1964}.
We integrate each YSC for a maximum time $t=100\,\mathrm{Myr}$. 

As a  result of the Kroupa IMF and of the formalism we adopted for BH formation, the total number of BHs per each star cluster follows a linear relation  $n_{\rm{BH}} \, = \, \mathcal{A}\, + \,\mathcal{B} \,\,(m_{\rm{SC}}/{\rm M}_\odot) \, $, where $ \mathcal{A} \, =\, 0.320\pm 8\times{}10^{-3}$ and $ \mathcal{B} \, = \, 0.003\,\pm 1.5\times{}10^{-5} $, respectively. 
On average, the smallest YSCs ($m_{\rm SC}\approx 300$ \Ms) produce $1$ BH, while the most massive systems we simulated ($m_{\rm SC}\approx 1000$ \Ms) produce up to 3 BHs. Hence, we can form just one BBH in each YSC, on average. This is another crucial difference with respect to more massive star clusters.

Throughout this paper, we compare our results with those described in \cite{dicarlo2019a} and \cite{dicarlo2020}, who used the same code to study high-mass YSCs ($m_{\rm{SC}} \in [10^3,3\times{}10^4]$~M$_\odot$). 

 \cite{dicarlo2020} explore two density regimes: i) dense YSCs, following the \cite{marks2012} relation, with initial half-mass density $\rho_{\rm h}\ge{}3.4\times10^4$ M$_\odot$ pc$^{-3}$ (set~A),
 and ii) loose YSCs, 
 with initial half-mass radius $r_{\rm h}=1.5$ pc and initial half-mass density $\rho_{\rm h}\ge{}1.5\times10^2$ M$_\odot$ pc$^{-3}$ (set~B). Our YSCs are the extension of set~A of \cite{dicarlo2020} to low-mass systems.

Moreover, we compare our results to a set of isolated binaries 
 simulated with \mob{}. This sample is composed of
$3\times{}10^7$ binaries ($10^7$ for each metallicity).
The primary masses of the isolated binaries are drawn from a Kroupa \citep{kroupa2001} mass
function between $5$ and $150$ M$_\odot$. The initial orbital properties of the isolated binaries follow the same distributions as we described for original binaries in YSCs.

\section{Results}
\label{results}


\begin{figure*}
\centering
\includegraphics[width=0.8\textwidth]{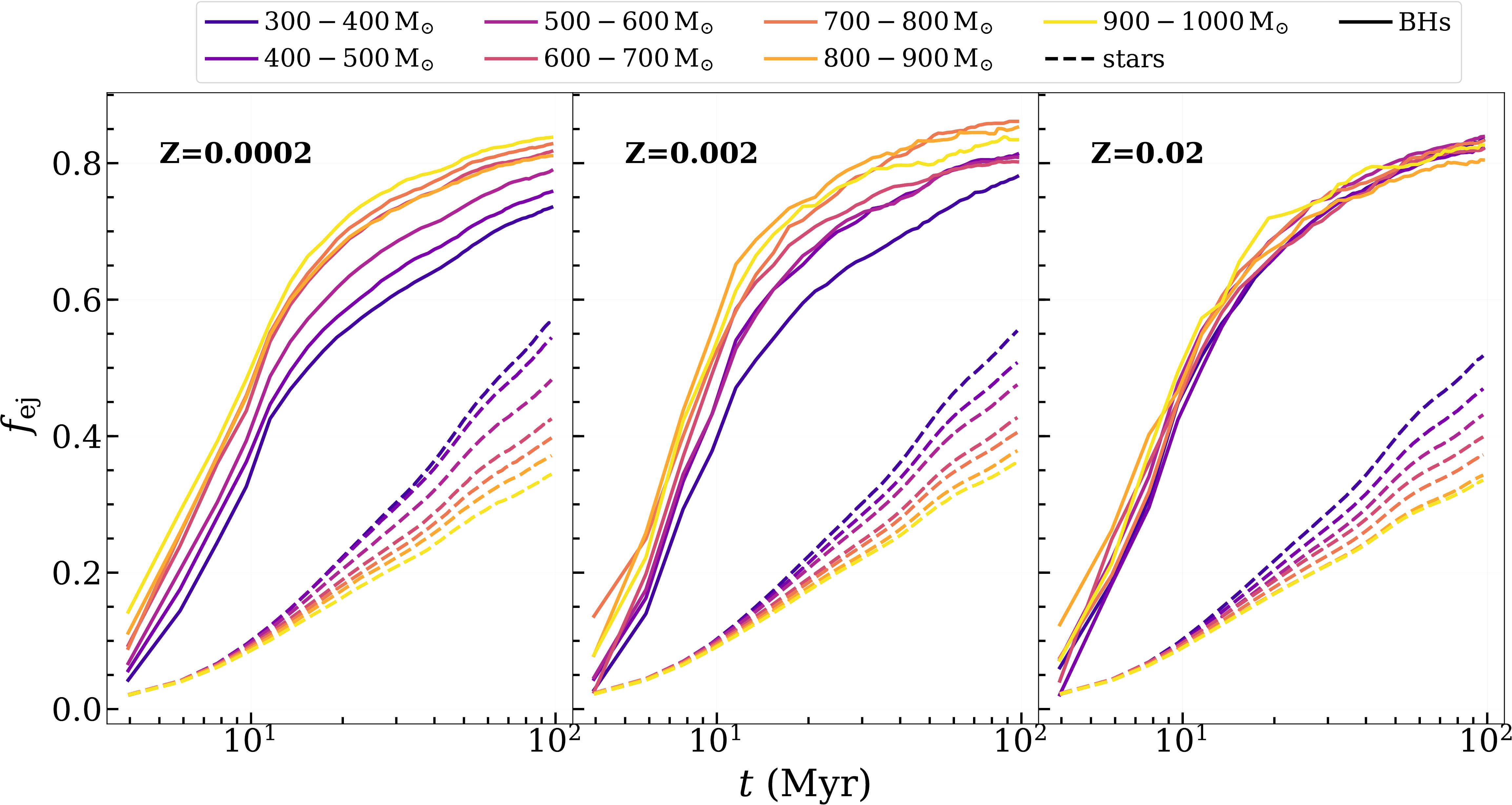}
\caption{ Solid (dashed) lines: fraction of ejected BHs (stars)  with respect to the total number of  BHs (stars). 
From left to right: $Z=0.0002$, 0.002 and 0.02. We adopt the same colour palette as in Figure~\ref{fig:9plot}.}
\label{fig:rbh}
\end{figure*}

\subsection{YSC Evolution} 
\label{ysc_ev} 

Figure~\ref{fig:9plot} shows the global evolution of YSCs' main physical
parameters: half mass radius $r_{\rm{h}}$, core radius $r_{\rm{c}}$ \citep{Casertano} and 
stellar density $\rho_{\rm{h}}$ at the half-mass radius.
 Heavy YSCs ($700<\,m_{\rm SC}/{\rm M}_\odot{}\,<1000$ \Ms{}) expand more slowly
than lighter systems ($300<\,m_{\rm SC}/{\rm M}_\odot{}\,<700$), and the expansion is
generally steeper in metal-poor than in metal-rich systems. 
In the initial conditions, all YSCs are tidally under-filling. Then, all the clusters
expand and the smallest ones (in the mass range $300-700$~\Ms{}) reach and exceed
 the tidal radius, 
becoming tidally
over-filling. Only the most massive systems ($700 <m_{\rm SC}/{\rm M}_\odot{}<
1000$) remain tidally under-filling for the entire simulation. 
Hence, heavy clusters survive longer against the expansion, since they 
 resist more efficiently the external tidal field. 

As anticipated in Section \ref{intro}, low-mass YSCs have extremely short 
relaxation time scale $t_{\rm rlx}$, of the order of a few tens of Myr, and undergo core collapse over an even shorter time scale ($t_{\rm cc}\approx{}0.2\,{}t_{\rm rlx}$, \citealt{fujii2014}). 
 Our star clusters are essentially generated as already
core-collapsed because of the high initial density 
of our models \citep{marks2012}. 
The evolution of the core radius is thus a rapid expansion, which
 is rather faster in lighter systems. After  $\sim{}100$ Myr,  
$r_{\rm c}$ has grown by a factor of $\sim{}28$ ($16$) if $m_{\rm SC}=300$ M$_\odot$ ($m_{\rm SC}=1000$ M$_\odot$). The global
evolution of the core radius is similar for each metallicity.

The initial rapid expansion of the clusters coincides
with an initial drop of the density, which 
is initially $\rho{}_{\rm h}\sim{}10^4$ \Ms~pc$^{-3}$ and decreases to $\rho{}_{\rm h}\sim 10^{-2}$ ~\Ms~pc$^{-3}$ ($\rho{}_{\rm h}\sim 1$~\Ms~pc$^{-3}$) at 100 Myr if
$m_{\rm SC}=300$~M$_\odot$ ($m_{\rm SC}=1000$~M$_\odot$).

\subsection{Dynamical Ejection of BHs}

Figure~\ref{fig:rbh} shows the evolution of the fraction $f_{\rm ej}$ of ejected BHs with respect to all the BHs formed in each star cluster. For comparison, we also show the fraction of ejected stars over the whole cluster population. 
The fraction of ejected BHs increases with time, initially very fast and then, after
$\approx 10$ Myr, with a more gentle slope. The change of the slope approximately corresponds to the time at which all the BHs have formed and to the time at which the density has dropped by $\sim{}3$ orders of magnitude (Figure~\ref{fig:9plot}).  
To draw Figure~\ref{fig:rbh}, we have considered only BHs after their formation: we have not included their progenitors. For this reason, the time shown in the $x-$axis should be regarded as the maximum between the time of ejection and the time of BH formation: several massive stars are ejected before they collapse to a BH. 

Between $72$ and $85$\% of all BHs are ejected from the star cluster by the end of the simulation. Moreover, even those BHs that remain inside the YSC spend most of their life in a low-density environment, given the drop of the density after $\sim{}10$ Myr. Hence, the BHs formed in our YSCs undergo strong dynamical encounters and exchanges only up to a few $\times{}10$ Myr, while BHs formed in globular clusters can undergo a long chain of exchanges and three-body encounters before being ejected. 

If we look at the metal-poor star clusters ($Z=0.0002,$ $0.002$), the fraction of ejected BHs is higher in the heavier YSCs than in the lighter
ones, because the higher density of the heavier YSCs triggers more interactions and increases the
chance to eject BHs via Spitzer instability \citep{spi87}. 

This difference between lighter and heavier YSCs tends to disappear at solar metallicity, where the majority of BHs are relatively low-mass ($\sim{}5-10$ \Ms{}).

The fraction of ejected stars shows significant differences with respect to the one of BHs: after the first $10$ Myr, the slope of the fraction of ejected stars becomes even steeper, while the fraction of ejected BHs becomes flatter. At the end of the simulations $\sim{}30-55$\% of the stars are ejected from their parent YSC. The lighter YSCs lose a higher fraction of stars than the heavier ones: at the end of the simulation, $f_{\rm ej}\sim{}0.5-0.55$ and $\sim{}0.30-0.35$ in the YSCs with $m_{\rm SC}\in[300,400]$ \Ms{} and $m_{\rm SC}\in[900,1000]$ \Ms{}, respectively, with a very mild dependence on metallicity. This difference between BHs and stars reflects the different mechanism leading to their ejection. BHs are more massive than the average stellar mass, efficiently sink to the core by dynamical friction and are ejected by close dynamical encounters. In contrast, most of the ejected stars are light objects, populate the outskirts of their parent YSC and are tidally removed from it. The fraction of ejected stars is thus larger in the smaller YSCs, because these over-fill their tidal radius by the end of the simulation.

\subsection{Exchanged versus Original BBHs}

We label as \emph{exchanged BBHs} those BBHs that form through dynamical exchanges or other kind of encounters, and we call  \emph{original BBHs} those
 BBHs that descend from the evolution of an original binary star
(i.e. their stellar progenitors
were already bound in the initial conditions). 
With the term \emph{dynamical BBHs}, we indicate all the BBHs that evolve in YSCs, both original and exchanged. 
In fact, YSC dynamics affects not only exchanged binaries, but even original binaries: close dynamical encounters shrink (or widen) the semi-major axis of a binary system, change its eccentricity and can even break it. In general, lighter and wider binaries (soft binaries) tend to be widened/ionized, while massive and tight binaries (hard binaries) tend to increase their binding energy and shrink their orbit by close encounters \citep{heggie1975}.

\begin{figure}
\centering
\includegraphics[width=0.45\textwidth]{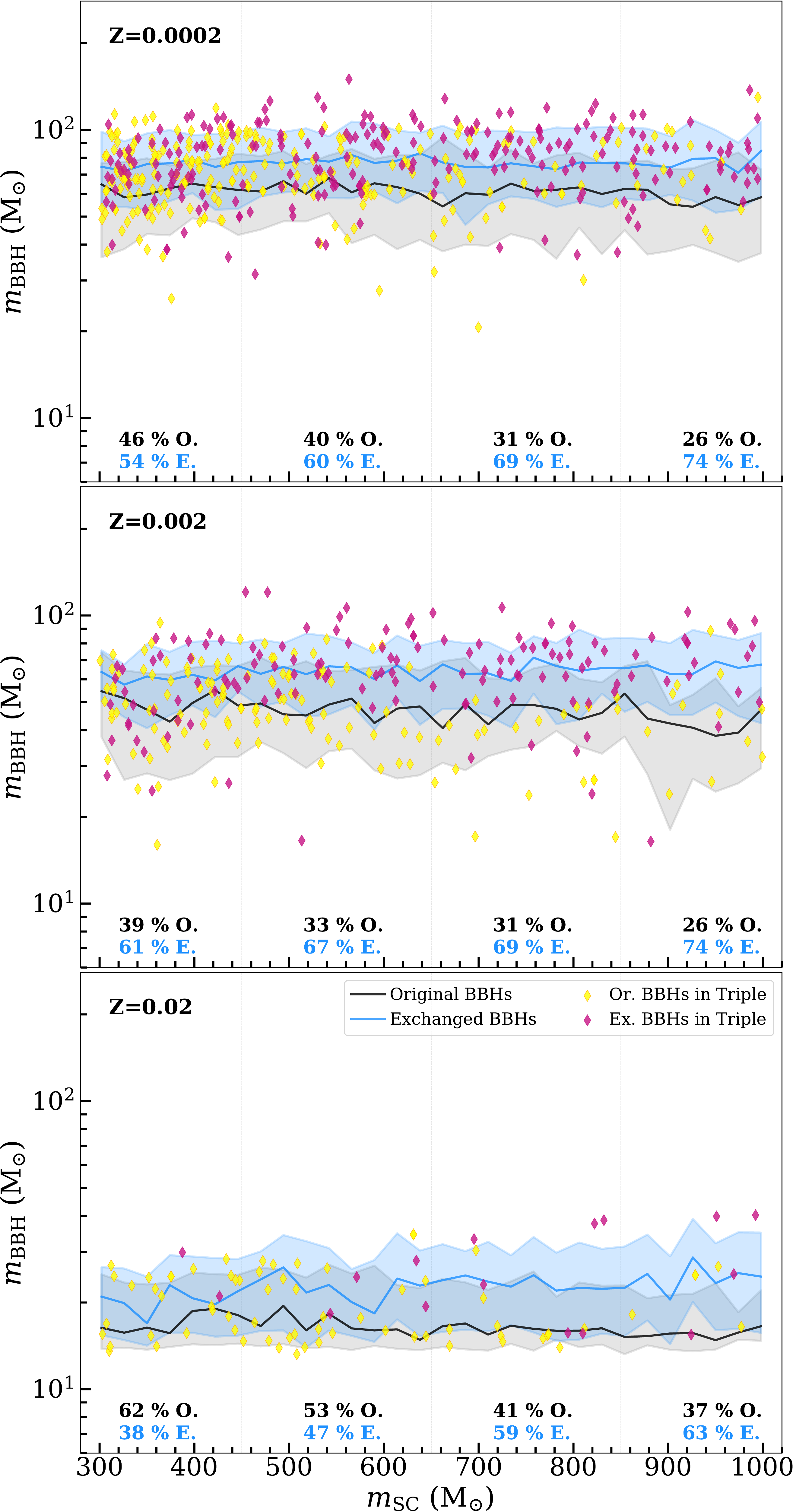}
\caption{ BBH total mass ($m_{\rm{BBH}}$) as a function of the YSC mass ($m_{\rm{SC}}$)
at $Z=0.0002$, $0.002$, $0.02$, (upper, middle and lower panel, respectively) at the end of the simulations ($100$ Myr).
Solid lines indicate the median value of the BBH mass: blue for exchanged BBHs and black
for original BBHs. The corresponding shaded areas (light blue and grey) indicate the
confidence interval at the $20$th and $80$th percentile. Yellow and magenta diamonds indicate original and exchanged
BBHs in triple systems, respectively. 
 At each metallicity, we divide the YSCs into four bins of mass: $m_{\rm{SC}}\in[300,450]\,{}{\rm M}_\odot$, $m_{\rm SC}\in[450,650]\,{}{\rm M}_\odot$, $m_{\rm SC}\in[650,850]\,{}{\rm M}_\odot$, and
$m_{\rm{SC}}\in[850,1000]\,{}{\rm M}_\odot$. The percentage of original (O.) and exchanged (E.) BBHs per each of these bins is reported in each panel. }
\label{fig:mtot_bbh_mysc}
\end{figure}

Figure \ref{fig:mtot_bbh_mysc} shows the distribution of BBH masses 
  as a function of star cluster mass and for different metallicities. The mass of both original and exchanged BBHs is mainly affected by the metallicity of the progenitors and does not depend on the mass of the host star cluster.

Exchanged BBHs are, on average, more massive than original BBHs:  the median value of the mass of exchanged BBHs is $\sim 80$, $\sim 60$ and $\sim 20$ \Ms{} at metallicity 
$Z=0.0002$, $0.002$ and $0.02$, respectively.
In contrast, original BBHs are, on average, less massive
( $\sim 60$, $\sim 30$ and $\sim 10$ \Ms{} at $Z=0.0002$, $0.002$ and $0.02$, respectively).

The percentage of original (exchanged) BBHs significantly decreases (increases) as the cluster mass increases.
In fact, heavier clusters allow a larger number of  dynamical encounters that tend to disrupt original binaries and to form more BBHs via exchanges.

BBHs in triple systems (also shown in Figure~\ref{fig:mtot_bbh_mysc}) form more frequently in metal-poor environments:
55 \%  (33\%) of all BBHs are members of triple systems at $Z=0.0002$ ($Z=0.002$).
At solar metallicity, only 10\% of BBHs are in triples.
The ratio between original and exchanged triples\footnote{In this case, we
refer to the inner binary of the triple which may be an original or an exchanged
BBH.} is quite balanced in metal-poor systems:
at $Z=0.0002$ ($Z=0.002$), $49\%$ ($48 \%$) of all BBHs in triples are exchanged systems. In contrast, at solar metallicity, the vast majority of BBHs in triples are original ($80 \%$).


\begin{figure*}
\centering
\includegraphics[width=0.9\textwidth]{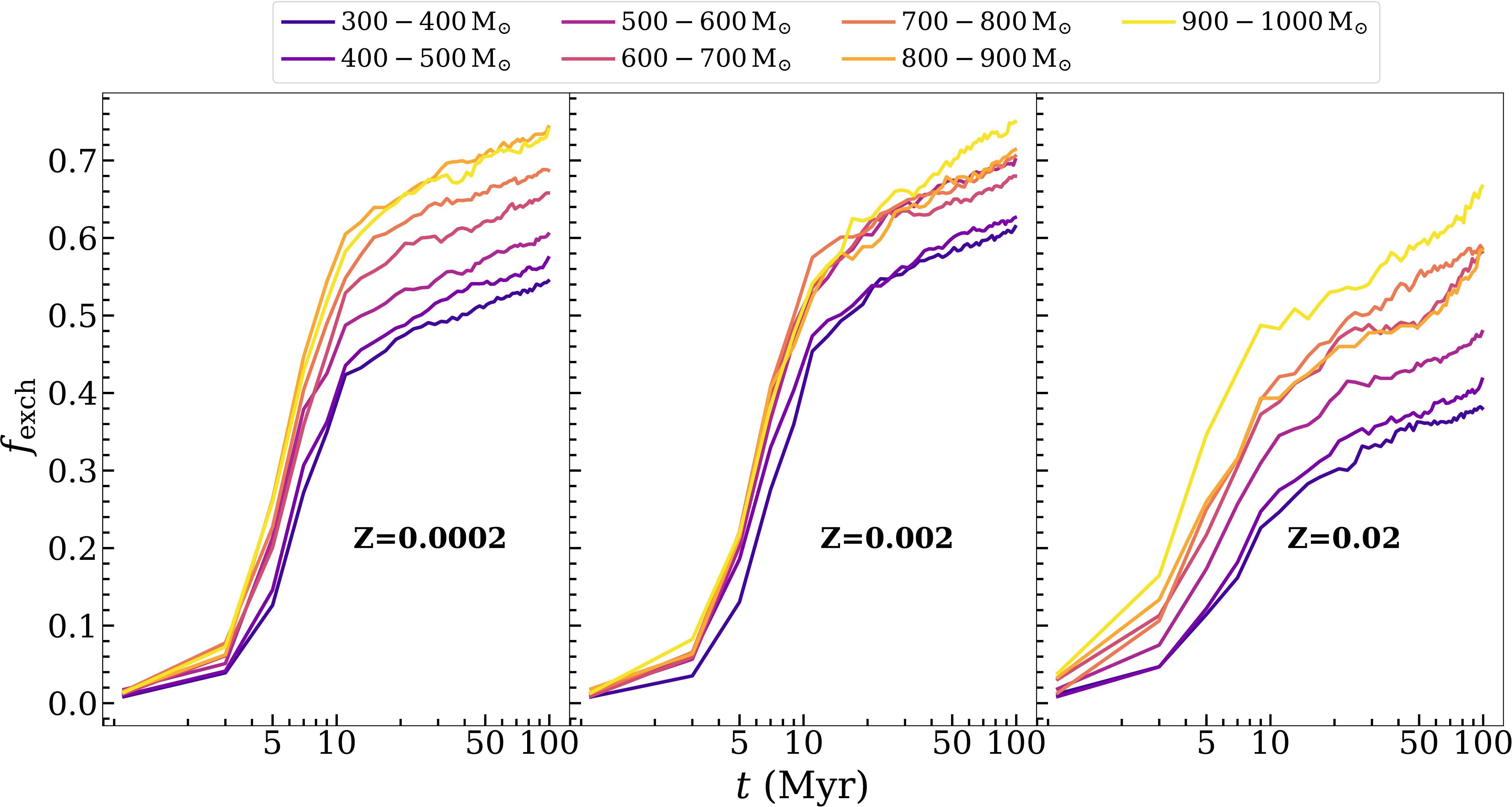}
\caption{ Evolution of the fraction of exchanged BBHs with respect to all BBHs at $Z = 0.0002$ (left), $Z = 0.002$ (middle) and $Z = 0.02$ (right). The colour map is the same as in Figure \ref{fig:9plot}.  }
\label{fig:frac}
\end{figure*}

Figure~\ref{fig:frac} shows the temporal evolution of the fraction of
exchanged BBHs
\begin{equation}
f_{\rm{exch}}(t) = \frac{N_{\rm exch}(t)}{N_{\rm exch}(100\,{}{\rm Myr})+ N_{\rm orig}(100\,{}{\rm Myr})},
\end{equation}
where $N_{\rm exch}(t)$ is the number of exchanged BBHs (or BBH progenitors) at each time step ($t$), while $N_{\rm exch}(100\,{}{\rm Myr})$ and $N_{\rm orig}(100\,{}{\rm Myr})$ are the number of exchanged and original BBHs at the end of our simulations ($t=100$ Myr). When calculating $N_{\rm exch}(t)$ we consider not only BBHs that are already formed at time $t$, but also exchanges involving their stellar progenitors. 

The fraction of exchanged BBHs has a fast growth in the first $\sim{}10$ Myr and a shallower one in the next stages. This happens because exchanges are more efficient when the density is higher and figure~\ref{fig:9plot} shows that the density drops at $\approx{}10$ Myr. Moreover, $f_{\rm exch}$ strongly depends on both YSC mass and metallicity. The heavier mass clusters are obviously more efficient in forming exchanged systems, because of their higher density and longer lifetime. For example, at $Z=0.0002$, $f_{\rm exch}\approx{}0.55$ ($\approx{}0.75$) in YSCs with mass $300-400$ M$_\odot$ ($900-1000$ M$_\odot$) at the end of the simulations. 

Metal-poor YSCs ($Z=0.0002,$ $0.002$) have higher values of $f_{\rm exch}$ than metal-rich ones. For example, at the end of the simulation, $f_{\rm exch}\approx{}0.55$ ($\approx{}0.35$) in YSCs with mass $300-400$ M$_\odot$ at $Z=0.0002$ ($Z=0.02$). This difference springs from the dependence of BH mass on progenitor's metallicity: dynamical exchanges favour the formation of massive binaries, which are more energetically stable \cite[e.g.,][]{hills1980}; hence, massive BHs in metal-poor YSCs are more efficient in pairing up via exchanges.


\begin{figure}
\centering
\includegraphics[width=0.45\textwidth]{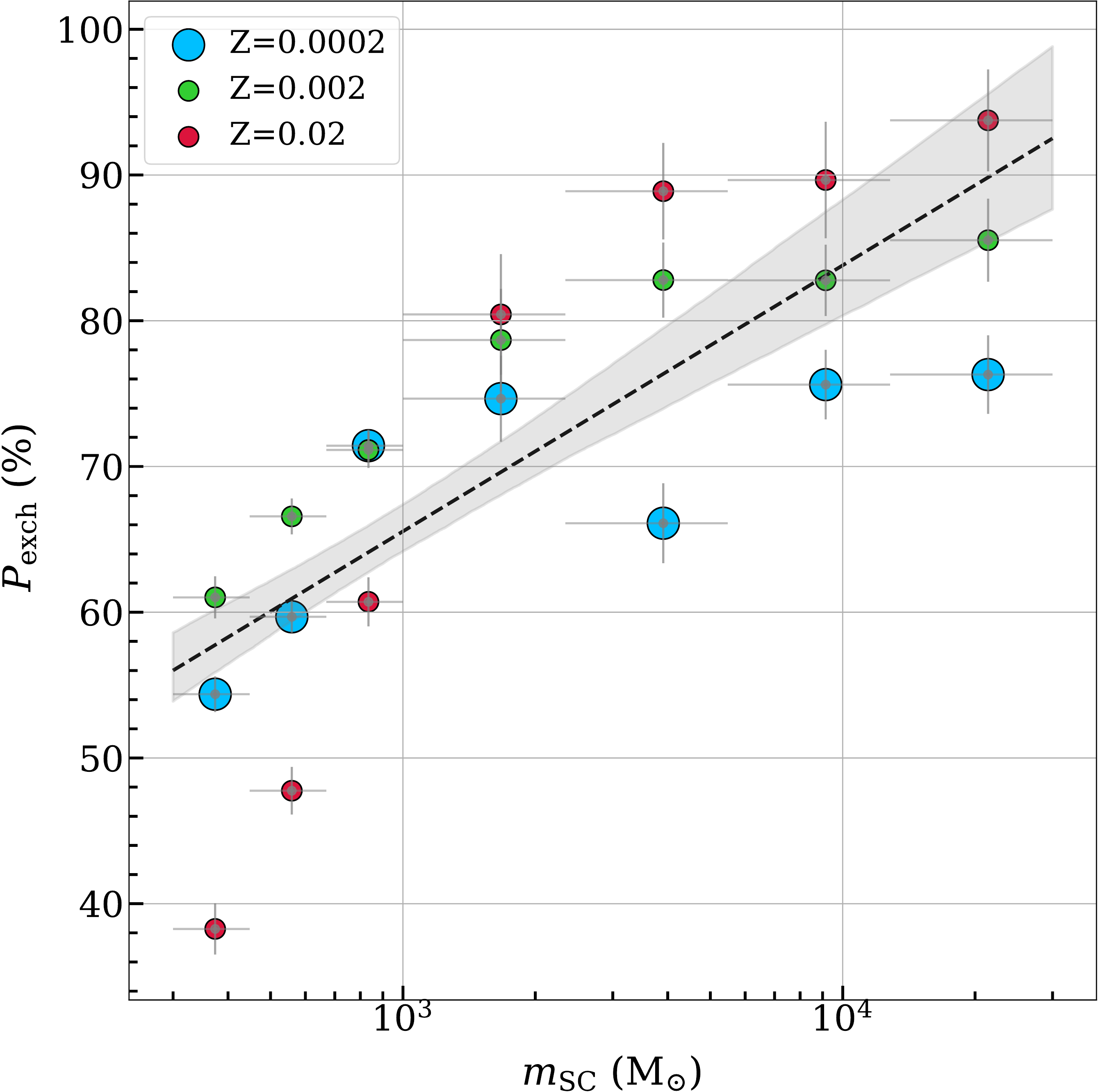}
\caption{ Percentage of exchanged BBHs ($P_{\rm exch}$) per cluster mass bin ($m_{\rm SC}$) in low-mass and high-mass YSCs. Blue, green and red points refer to $Z=0.0002$, $0.002$ and $0.02$, respectively. The error bars are Poissonian uncertainties. 
The black dashed line indicates the best linear fit (in the lin$-$log plane) and the grey bands refer to the
confidence interval at the $20$th and $80$th percentile. 
The YSCs with $m_{\rm SC}<1000$ M$_\odot$ are from this work, while the YSCs with $m_{\rm SC}\ge{}1000$ M$_\odot$ are from \citet{dicarlo2020}, set~A.}
\label{fig:perc_exch_fit}
\end{figure}

Figure~\ref{fig:perc_exch_fit} shows the percentage of exchanged BBHs at the end of the simulation ($P_{\rm{exch}}$) as a function of the star cluster mass $m_{\rm SC}$. In this Figure, we include the YSCs from set~A of \cite{dicarlo2020}, which span from $m_{\rm SC}=10^3$ M$_\odot$ to $3\times{}10^4$ M$_\odot$. $P_{\rm exch}$ increases with the mass of the YSC.  The percentage of exchanged BBHs increases from $\approx 60$ \% in low-mass YSCs up to more than $\approx 90$ \% in high-mass YSCs. 
Fitting the points in Fig.~\ref{fig:perc_exch_fit} with the linear fit $P_{\rm exch}=a \,+\, b \,{} \log_{10}{(m_{\rm SC}/1000\, M_{\odot})}$,  
we obtain parameters $a = 65.1 \pm 3.1 $ and $b =18.9 \pm 0.2$. We also  ran a Pearson’s correlation test, to estimate the significance of the $P_{\rm{exch}}-\log_{10} m_{\rm SC}$ correlation.  We obtain a Pearson's coefficient $r>0.5$, implying a rather strong correlation between $P_{\rm exch}$ and $m_{\rm SC}$. This correlation is steeper for metal-rich ($Z=0.02$) than for metal-poor clusters ($Z=0.0002$), possibly because in metal-poor environments original BBHs are more massive and more difficult to break by dynamical exchanges.

Massive YSCs, as those studied in \cite{dicarlo2020}, have a larger probability to produce exchanged binaries because of the higher rate of dynamical encounters, which tend to destroy original binaries. In contrast, because of the fast dynamical evolution and the rapid central density decrease, low-mass YSCs allow the survival of more original BBHs, while the formation of new binaries trough dynamical exchanges is  suppressed with respect to more massive clusters.  Moreover, high-mass YSCs host more BHs ($\approx$30 if $m_{\rm SC}\approx{}10^4$ M$_\odot$) than low-mass systems: a star cluster with $m_{\rm SC}\approx{}10^3$ M$_\odot$  hosts, on average, 3 BHs  (see Section~\ref{sec:init_cond}).  Hence, a low-mass YSC may host at most one BBH plus, in a few cases, a single BH.  
Since massive stars tend to be part of binary systems with other massive stars (as a consequence of the algorithm we used to generate the initial conditions, which is supported by observations, see Section~\ref{sec:init_cond}), in the lowest mass clusters  the only BBH tends to be the result of the evolution of the original binary composed of the two most massive stars. When there are no other BHs in the star cluster, it is extremely difficult that such a massive original binary undergoes an exchange, because the probability of an exchange is strongly suppressed if the intruders are less massive than the two members of the binary system \citep{hills1980}. For the same reason, low-mass YSCs favour the formation and merger of BH-neutron star binaries \citep{rastello2020}, which are, instead,  highly suppressed in more massive star clusters \citep{ye2020}.

\subsection{Isolated vs Dynamical BBHs}


\begin{figure}
\centering
\includegraphics[width=0.45\textwidth]{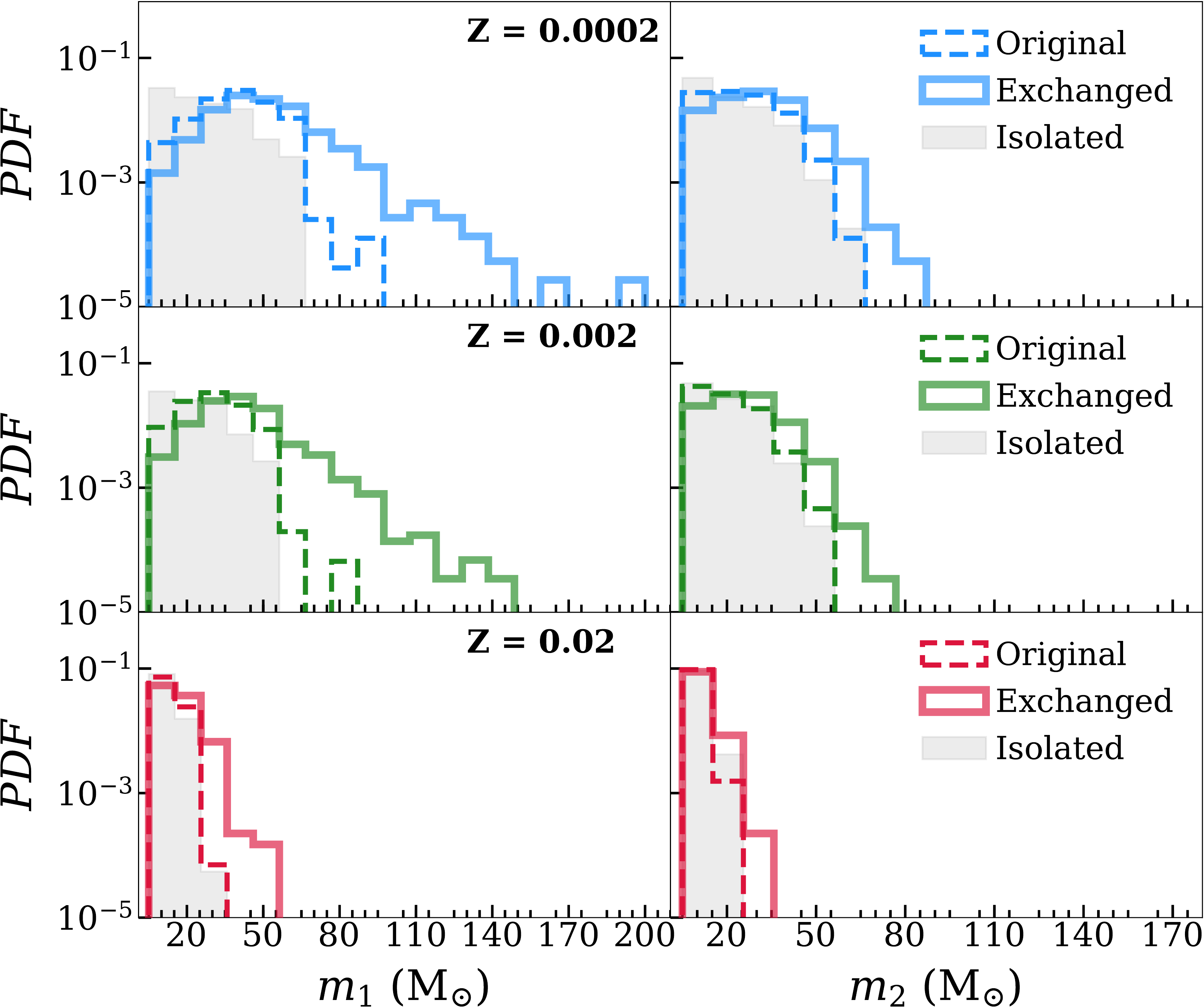}
\caption{ Mass of primary ($m_{1}$, left panels) and secondary ($m_{2}$, right panels) BHs of formed BBHs in YSCs and
in isolation (filled grey histograms). The panels from top to bottom refer
to the three metallicities: Z = $0.0002$ (blue histogram),
Z = $0.002$ (green histogram) and Z = $0.02$ (red histogram), respectively.
 Solid lines: exchanged BBHs; dashed lines: original BBHs.}
\label{fig:m1_m2_tot}
\end{figure}

We compare our dynamical BBHs with a set of \emph{isolated BBHs}, that formed in the field from isolated binary evolution and that we evolved with {\sc mobse} without dynamics. Figure~\ref{fig:m1_m2_tot} compares the mass distribution of BBHs
formed in YSCs with that of BBHs formed in isolation. This Figure shows both BBHs that merge within a Hubble time (hereafter, merging BBHs) and BBHs that are too loose to merge over the age of the Universe. The maximum mass of the primary components of exchanged BBHs, $m_{\rm 1,\,{}max}$, is significantly higher than that of original and isolated binaries, especially
in metal-poor environments: at $Z=0.0002$, $m_{\rm 1,{}max}\approx{}197$, $\approx{}91$ and $\sim{}63$ M$_\odot$ for exchanged, original and isolated BBHs, respectively. The most massive BHs in exchanged binaries originate from the merger  of two (or more) stars, which then pair up dynamically \citep{portegieszwart2004}. Some of these masses are in the pair-instability mass gap \citep{dicarlo2020b}. 
Light BBHs ($m_1\lesssim{}25$ M$_\odot$) are more common in isolated BBHs than in both original and exchanged BBHs, especially at low metallicity.

\subsection{BBH Mergers}

We now focus on merging BBHs, i.e. BBHs that reach coalescence within a Hubble time. In our dynamical simulations, we found $\sim{}430$ merging BBHs. 
 $60$\% of the merging BBHs are at metallicity $ Z = 0.0002$, $35$\% at $Z = 0.002$ and $5$\% at $Z = 0.02$.

 The percentage of original merging
BBHs is always larger than that of exchanged binaries: $\approx 87$\%, $\approx 84$\% and $\approx 95$\% of  merging BBHs are original at
$Z =0.0002$, $0.002$ and $0.02$, respectively.

Merging BBHs follow the same trend we have already seen in Figure~\ref{fig:perc_exch_fit} for all (merging and non merging) BBHs: $\approx 90$\% of the merging BBHs in low-mass YSCs ($300-10^3$ M$_\odot$) are
original binaries,  while in more massive systems \citep{dicarlo2020} the percentage of original merging BBHs drops to $\approx60$\%. However, when compared with Figure~\ref{fig:perc_exch_fit}, these percentages indicate that original BBHs are more efficient in merging than exchanged BBHs. In fact, exchanged BBHs are $\sim{}40-60$\% ($\sim{}80-90$\%) of all BBHs, but they represent only the $\sim{}10$\% ($\sim{}40$\%) of the merging BBHs in the least massive YSCs (most massive YSCs). The reason is that exchanged BBHs generally form with larger semi-major axis than original BBHs and cannot shrink efficiently via dynamical encounters, especially in the least massive YSCs.


\begin{figure}
\centering
\includegraphics[width=0.45\textwidth]{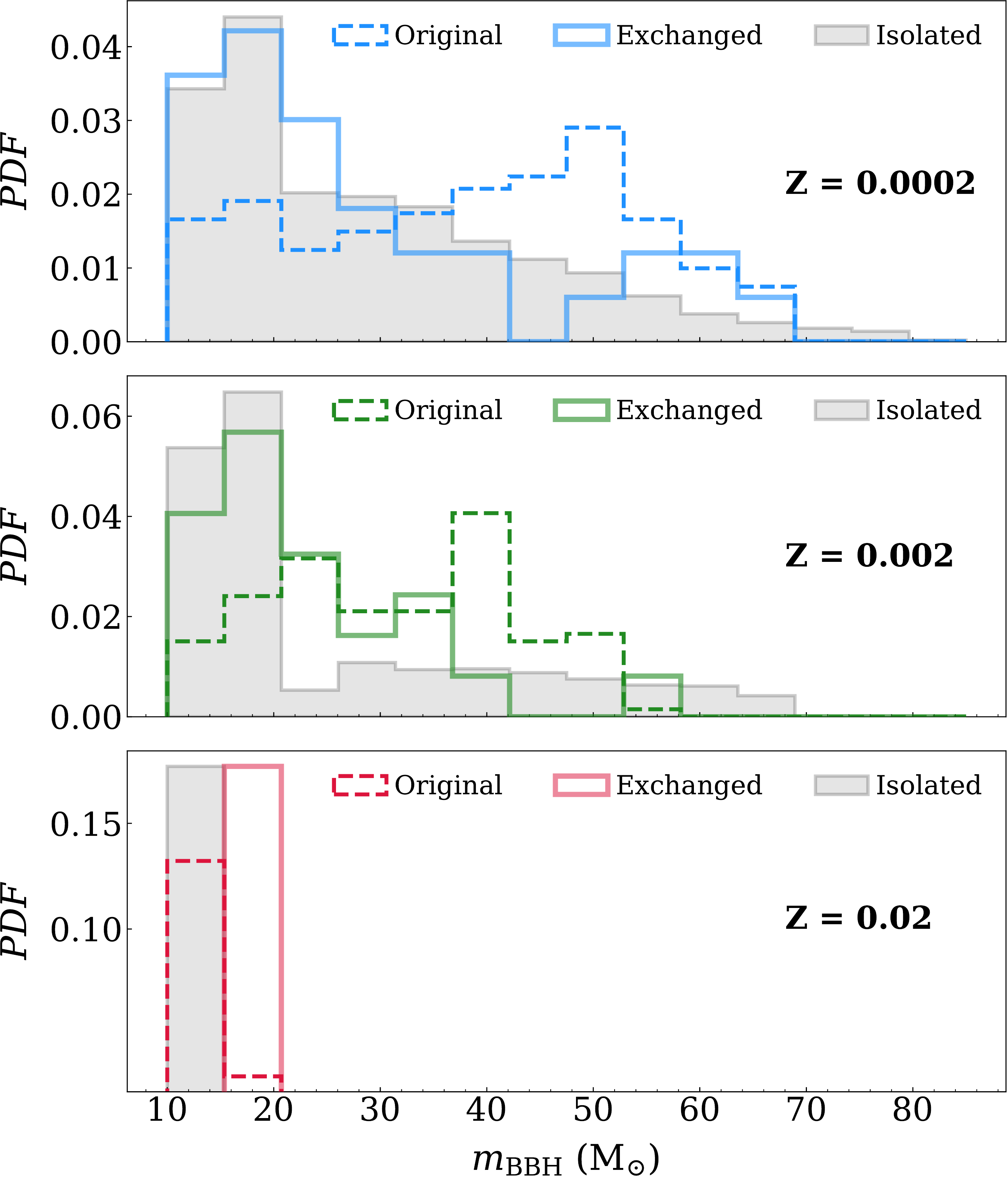}
\caption{ Distribution of total masses ($m_{\rm{BBH}}$ = $ m_{1} + m_{2}$) of merging BBHs.
Solid line: exchanged BBHs; dashed line: original BBHs; grey filled histogram: isolated BBHs.
The three plots display, from top to bottom: $Z = 0.0002$,
 $Z = 0.002$ and $Z = 0.02$.}
 \label{fig:mtotcoal}
\end{figure}

\begin{figure}
\centering
\includegraphics[width=0.45\textwidth]{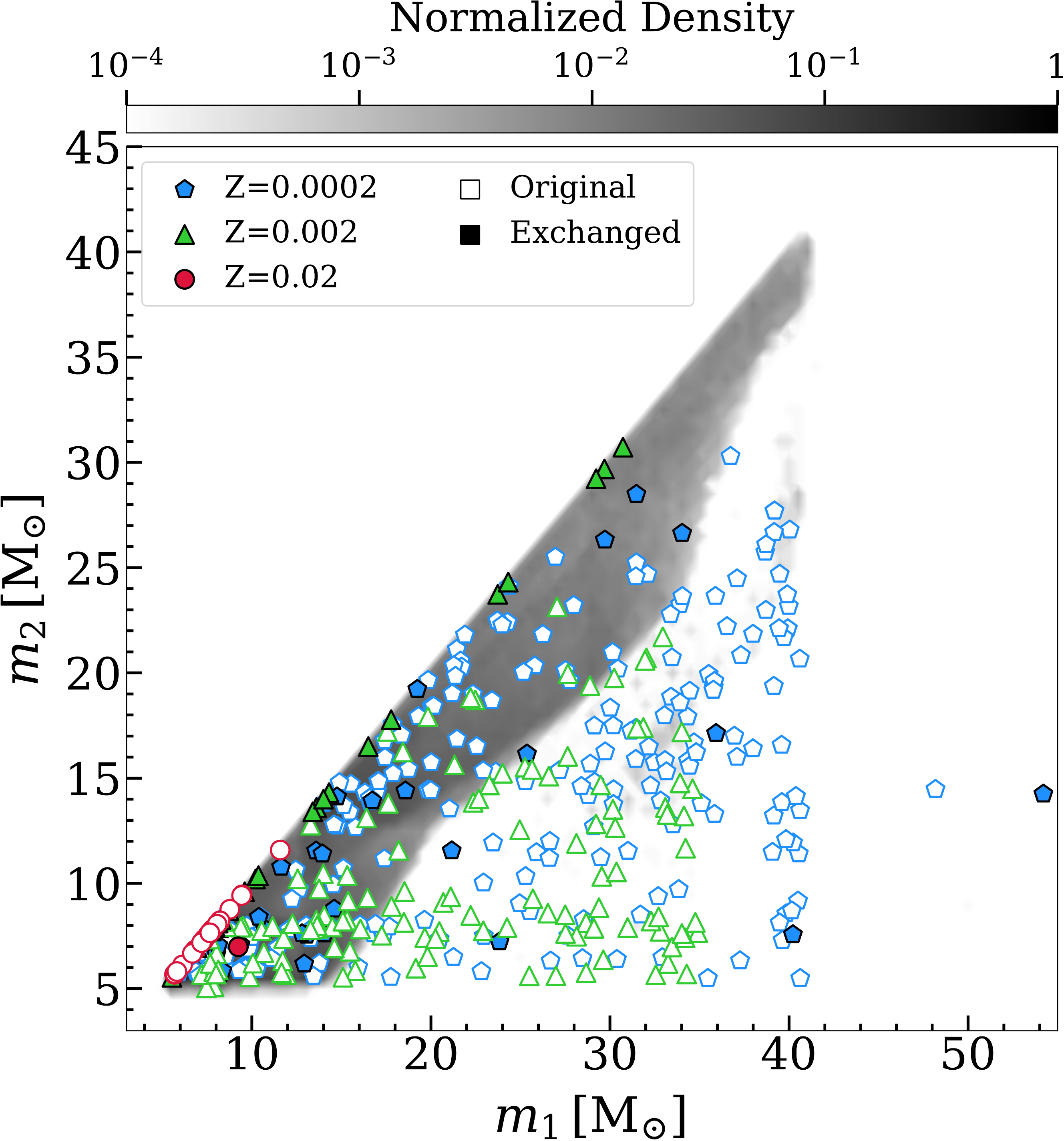}
\caption{ Mass of the primary ($m_1$) and secondary ($m_2$) component of merging BBHs in the low-mass YSCs presented in this work.
Empty symbols: original BBHs; Filled symbols: exchanged BBHs. Blue:
$Z= 0.0002$; green: 0.002; red: 0.02. Filled contours (grey colour map): isolated BBHs mergers for all three metallicities together, from our comparison sample.}
\label{fig:m1m2sara}
\end{figure}

\begin{figure}
\centering
\includegraphics[width=0.45\textwidth]{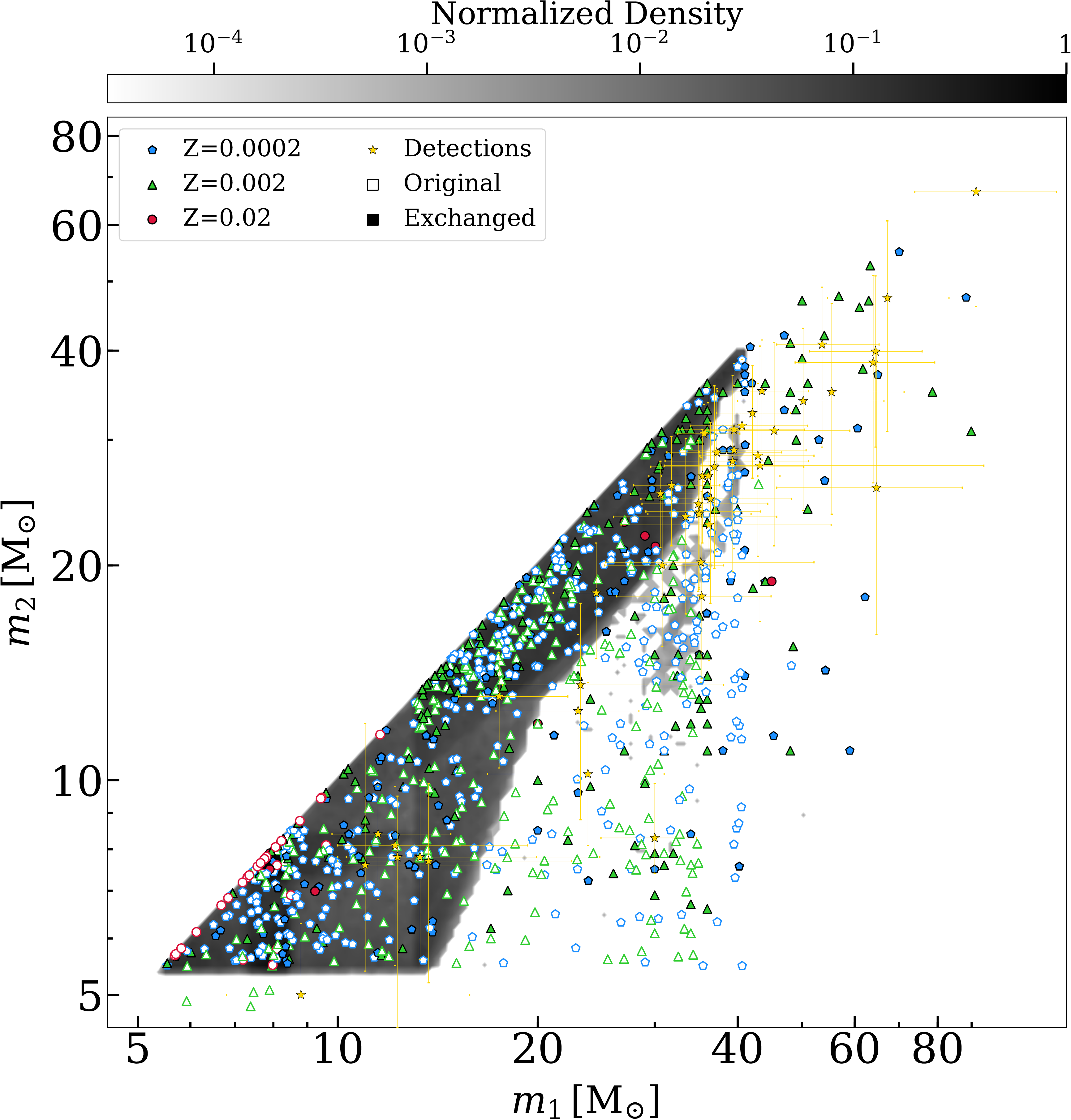}
\caption{ Same as Figure~\ref{fig:m1m2sara} but showing BBHs from both low-mass (this work) and high-mass YSCs (from \protect{\citealt{dicarlo2019a}} and \protect{\citealt{dicarlo2020}}). The yellow stars indicate all GWTC-2 BBHs \protect{\citep{abbottO3a}}. The error bars show the  90\%  credible interval. }
\label{fig:newtutti}
\end{figure}

\begin{figure}
\centering
\includegraphics[width=0.45\textwidth]{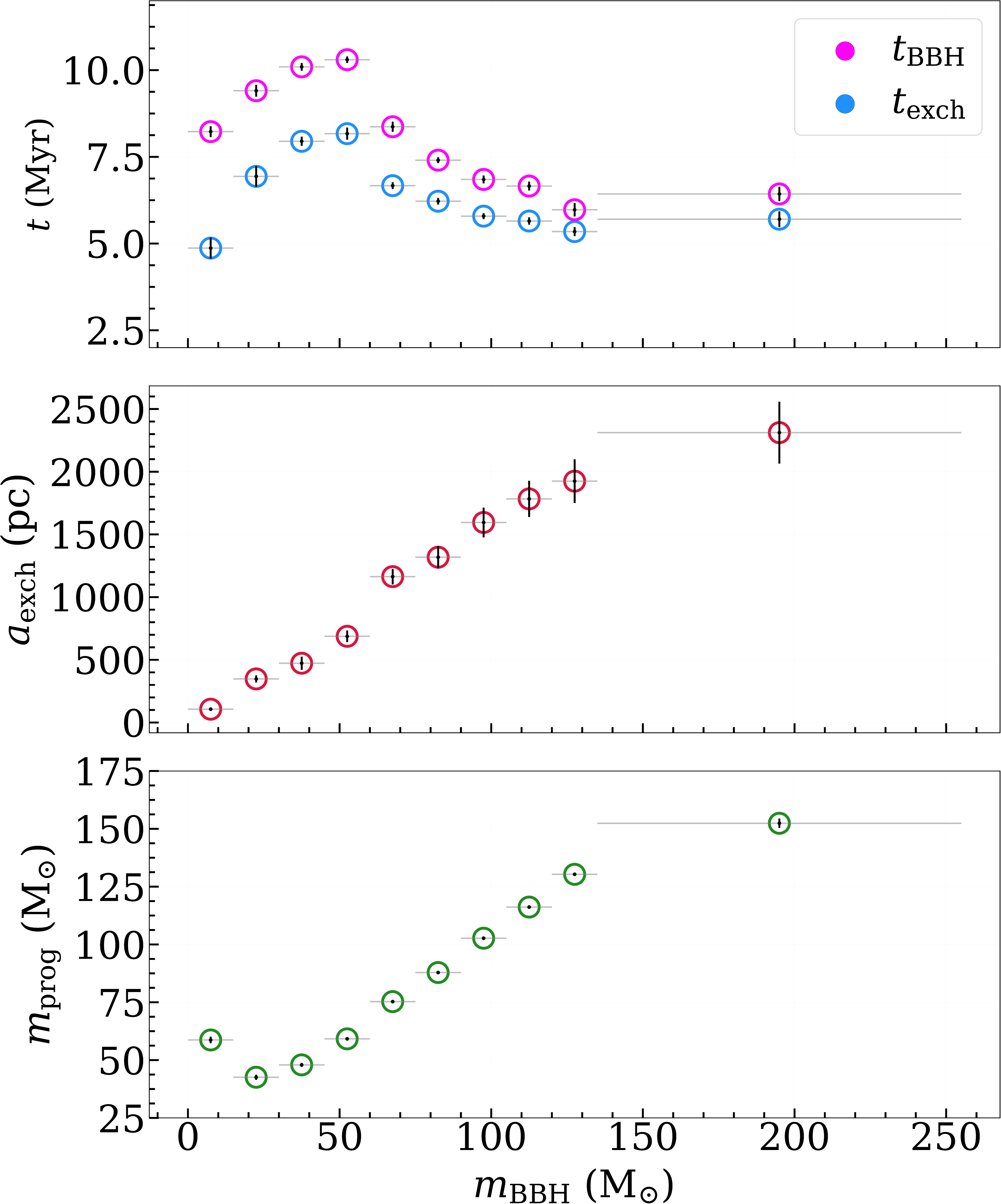}
\caption{Upper panel: time at which the first exchange occurs ($t_{\rm{exch}}$, blue circles) and time at which the second BH in exchanged BBHs forms ($t_{\rm{BBH}}$, violet circles) versus the total BBH mass ($m_{\rm{BBH}}$) for all the exchanged BBHs with $m_{\rm SC}\leq{}150$ \Ms{}. Central panel:  semi-major axis at time of the exchange ($a_{\rm exch}$) for all the exchanged BBHs. Lower panel: Total mass of BBH progenitor ($m_{\rm prog}$) for all exchanged binaries.
In all panels, the data point show the median value for all the exchanged BBHs in the same mass bin, while the error bars are obtained with the bootstrap method. BBHs with $m_{\rm SC}>150$ \Ms{} are shown in a unique large mass bin because of the low statistics in this mass range.
 } 
\label{fig:tex}
\end{figure}

Figure~\ref{fig:mtotcoal} shows the total mass ($m_{\rm BBH}\equiv{}m_1+m_2$) of merging BBHs in YSCs and in isolation, while Figure \ref{fig:m1m2sara} shows the mass of the primary ($m_{1}$)
 versus the mass of the secondary ($m_{2}$) BH component of merging BBHs in low-mass YSCs. Only two BBHs with primary mass $m_{1} > 45$ \Ms are found in metal poor systems ($Z = 0.0002$), one is original and the other is an exchanged binary. 
  
For comparison, Figure~\ref{fig:newtutti} shows the mass distribution of BBH mergers in low- ($300-10^3$ M$_\odot$) and high-mass YSCs ($10^3-3\times{}10^4$ M$_\odot$), combining the simulations presented in this work with those of 
\citet{dicarlo2019a} and \cite{dicarlo2020}. The maximum mass of merging BBHs is much higher in high-mass YSCs than in low-mass YSCs: the primary mass component is up to $\approx 90$ \Ms{} and $\approx54$ \Ms{} in high-mass and low-mass YSCs, respectively.  
All the systems with primary mass $m_1>50$ M$_\odot$ are exchanged BBHs.

Figures~\ref{fig:mtotcoal} and \ref{fig:m1m2sara} show that, even if we form exchanged BBHs with total mass up to $m_{\rm BBH}\sim{}230$ M$_\odot$ (Figure~\ref{fig:m1_m2_tot}) in the low-mass YSCs, only the least massive systems merge within a Hubble time ($m_{\rm BBH}<70$ M$_\odot$). Moreover, 
the exchanged BBH mergers have similar masses as the original ones. This is very different from what happens in more massive YSCs, where exchanged BBH mergers tend to be more massive than original BBH mergers (Figure~\ref{fig:newtutti}).

This fundamental difference springs from a number of effects. The first key ingredient is the time at which the dynamical exchange occurs. In low-mass YSCs (Figure \ref{fig:tex}), 
most of the exchanged BBHs pair up before the formation of the two BHs (when the binary is still composed of two stars, or of a star and a BH). 
The time of exchange ranges between $ 5.5 < t_{\rm{exch}} < 8$ Myr, while the BBH formation tends to occur later, $6.2 < t_{\rm{BBH}} < 10.2$ Myr. 
 
 The semi-major axis of the binary system right after the exchange $a_{\rm exch}$ (Figure~\ref{fig:tex}) plays a crucial role: if it is sufficiently small, the exchanged binary undergoes mass transfer and/or common envelope before the formation of the second BH. These processes have two effects. On the one hand, they further shrink the semi-major axis, leaving a binary system that may merge within a Hubble time. On the other hand, they remove a lot of mass from the progenitor stars, leading to the formation of smaller BHs \citep{linden2010,giacobbo2018b}. This is apparent from Figure~\ref{fig:tex}, where we see that the smallest BBHs ($m_{\rm BBH}\sim{}20$ M$_\odot$) have the smallest value of $a_{\rm exch}$ but are the result of 
 rather massive binaries.
 
 Finally, the most massive BH progenitors tend to form exchanged binaries with very large values of $a_{\rm exch}$. This happens because  such very massive binaries form via exchange and survive in the lowest-mass YSCs, even if they have a very large semi-major axis: they are hard binaries, even if $a_{\rm exch}\sim{}10^3$ AU. Even if they survive, their semi-major axis is so large that neither binary evolution processes nor further dynamical interactions (which become more and more rare, given the low density of the YSC, Figure~\ref{fig:9plot}) can shrink their orbit significantly. For a combination of these effects, exchanged BBH mergers tend to be relatively low mass in low-mass YSCs. In contrast, in more massive star clusters, even the most massive exchanged BBHs have a chance of dynamically shrinking their orbit, till they merge by gravitational wave decay.

\begin{figure}
\centering
\includegraphics[width=0.45\textwidth]{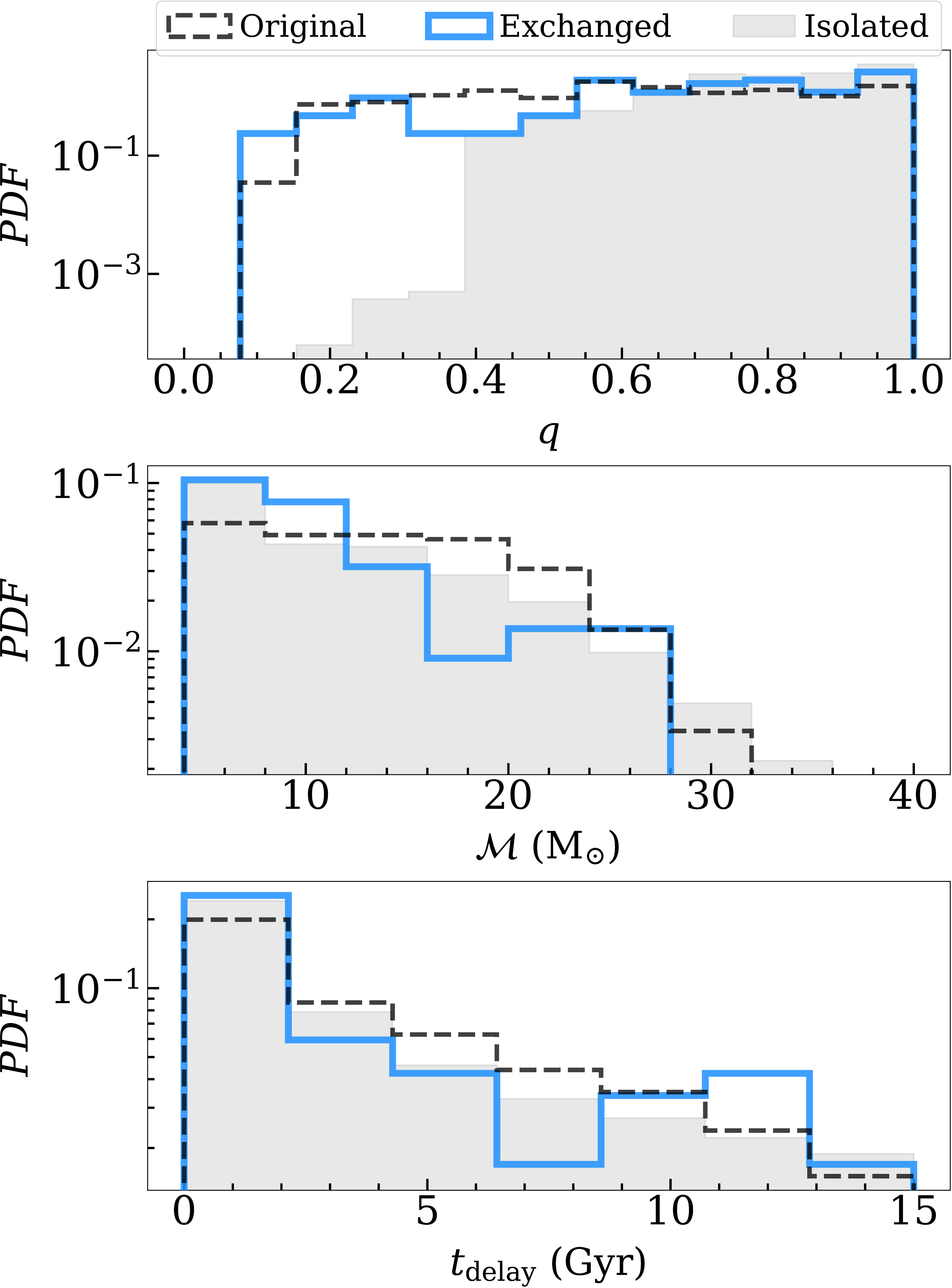}
\caption{
From top to bottom, distribution of mass ratio $q$,
 chirp mass $\mathcal{M}$ (\Ms{}) and delay time $t_{\rm{delay}}$ (Gyr) of merging BBHs in low-mass YSCs.
Black dashed line: original BBHs, blue solid line: exchanged BBHs.
Filled grey histograms refer to isolated BBHs. The three metallicities are displayed together.  }
\label{fig:info3coal}
\end{figure}

Figure~\ref{fig:info3coal} shows the distribution of the mass ratios
 $q=m_2/m_1$ of BBH mergers in low-mass YSCs.
Mass ratios of order of one are more common, but the distribution reaches a
minimum value $q\approx 0.13$.  Mass ratios $q<0.4$ are orders of magnitude more common in dynamical BBH mergers (both original and exchanged) than isolated binaries. This is an effect of dynamics that can lead to the pairing up of BHs with different masses, especially for exchanged binaries. The chirp mass of merging BBHs ($\mathcal{M}\,=\,{}(m_1\,{}m_2)^{3/5}\,{}(m1\,{}+\,{}m2)^{-1/5}$,  also shown in Figure~\ref{fig:info3coal}) ranges between a minimum of $\approx 4.7$ \Ms{}
to a maximum of $\approx 36$ \Ms{}. 
Similar to the distribution of total masses, also the one of chirp masses is skewed to lower values for the exchanged than for the original BBHs.
Finally, the distribution of delay times $t_{\rm{delay}}$ (Figure~ \ref{fig:info3coal})  of exchanged, original and isolated BBHs do not show particular differences.
The majority of coalescing BBHs merge in $\lesssim{} 2$ Gyr.
All the BBH mergers occur after the binaries have been ejected from the clusters, except for one case in which the BBH merges after $\approx30$ Myr, when is still bound to the cluster.

\subsection{BBH Survival Efficiency}

\begin{figure}
\centering
\includegraphics[width=0.45\textwidth]{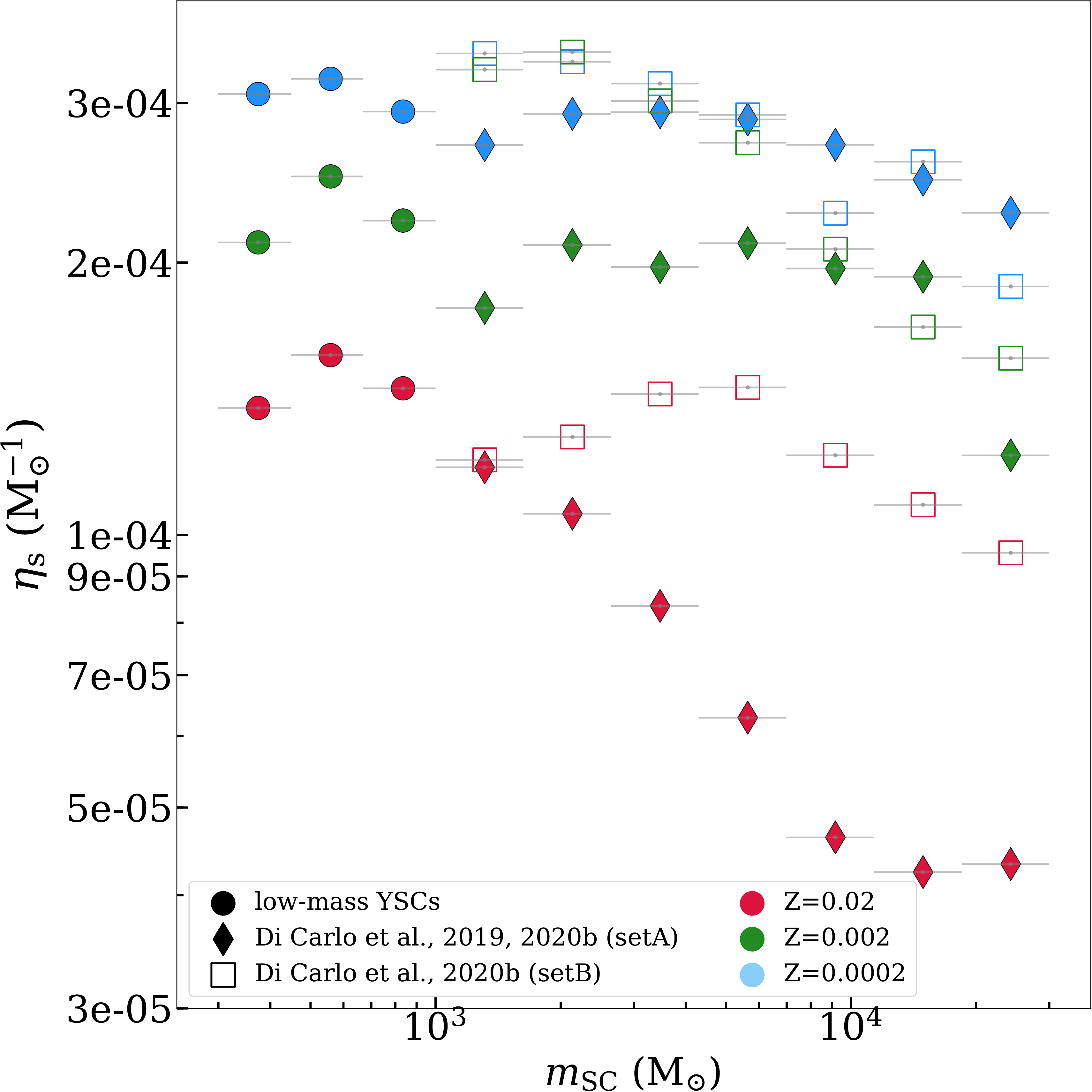}
\caption{BBH survival efficiency $\eta_{\rm s}$ as a function of YSC mass for $Z=0.0002$ (blue), $0.002$ (green) and $0.02$ (red).
Filled circles refer to the YSCs presented in this work ($m_{\rm SC}\in[300,\,{}1000]$ M$_\odot$), diamonds refer to set~A  of \citealt{dicarlo2020} and \citet{dicarlo2019a}, while empty squares refer to set~B of \citealt{dicarlo2020}. To derive $\eta_{\rm s}$ as a function of the cluster mass, we have divided the simulated YSCs in 10 log-spaced mass bins. The error bars are Poissonian uncertainties.  The size of each bin is indicated with grey lines along the x-axis.}

\label{fig:feff_bbh_bis}
\end{figure}

Figure~\ref{fig:feff_bbh_bis} shows the BBH survival efficiency ($\eta_{\rm s}$), defined as the number of BBHs that survive in the simulated YSCs at the end of the simulations ($N_{\rm BBH}$), divided by the total initial SC mass:
\begin{eqnarray}
\centering
\label{etas}
\eta_{\rm s}\,=\,\frac{N_{\rm BBH}} {M_\ast}.
\end{eqnarray}
We calculate $\eta_{\rm s}$ for both our YSCs ($m_{\rm SC}\in{}[300,\,{}1000]$ M$_\odot$) and those simulated by \citet{dicarlo2019a,dicarlo2020} ($m_{\rm SC}\in{}[1000,\,{}30000]$ M$_\odot$). We consider both BBHs that will merge in less than a Hubble time and looser BBH systems. 
The BBH survival efficiency is almost independent of mass in metal-poor YSCs ($Z=0.0002,$ 0.002), showing a weak decrease for large YSC mass. In contrast, metal-rich YSCs show an interesting behaviour: $\eta_{\rm s}$ drops by a factor of $\sim{}3$ going from low-mass to massive YSCs if we consider the densest star clusters (set~A of \citealt{dicarlo2020}), while it remains almost constant if we consider the looser star clusters (set~B of \citealt{dicarlo2020}). This indicates that, at solar metallicity, 
massive binary stars 
formed in dense stellar
systems (set~A) are more efficiently disrupted in the higher mass clusters. The same trend does not appear (or mildly appears) at lower metallicity and in looser star clusters (set~B). The reason why this behaviour appears only at high metallicity is that BHs produced at $Z=0.02$ are  lighter than those formed in metal-poor YSCs, facilitating the break up of binaries via ionization. In loose star clusters this effect is weaker, because dynamical encounters are less efficient. 
 For example, in massive metal-rich YSCs of set~A, no more than $10-15$\% of the original binaries survive till
the end of the simulation (see also Figure \ref{fig:perc_exch_fit}), while in set~B, the surviving binaries are $>35$\%.

\subsection{BBH Merger Efficiency versus Cluster Mass}


\begin{figure}
\centering
\includegraphics[width=0.45\textwidth]{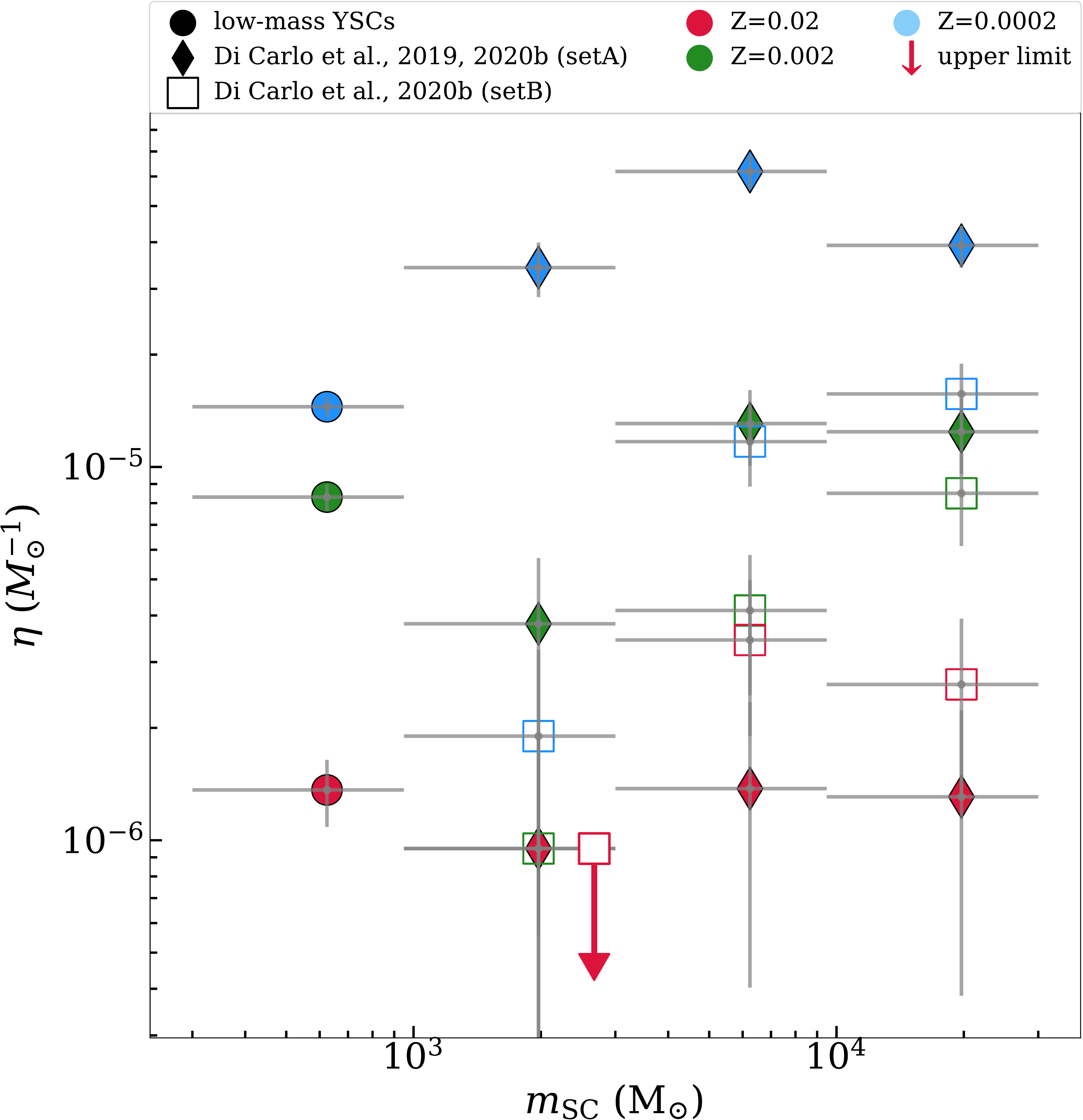}
\caption{ 
BBHs merger efficiency $\eta$  of YSCs. Metallicity $Z=0.0002$ (blue), $Z=0.002$ (green) and $Z=0.02$ (red).
Filled circles refer to our low-mass YSCs; filled diamonds refer to set~A of {\protect\cite{dicarlo2020}} and \citet{dicarlo2019a}, while empty squares refer to set~B of {\protect\cite{dicarlo2020}}.
The size of each bin is indicated with grey lines along the x-axis. The red arrow indicates the upper limit of $\eta$ in the first mass bin for set~B. The y-axis grey error bars are uncertainties from Poisson statistics.}
\label{fig:meffsc}
\end{figure}

The merger efficiency, $\eta{}$, is defined as the number of BBH mergers $\mathcal{N}_{\rm merg}$ within a Hubble time, divided by the total initial stellar mass of the simulated YSCs:
\begin{eqnarray}
\label{eq:eff}
\eta{} = \frac{\mathcal{N}_{\rm merg}}{M_{\ast}}
\end{eqnarray}
 where $M_\ast=\sum{}_i\,{}m_{\rm SC,\,{}i}$, where the index $i$ can run either over all the YSCs simulated at a given metallicity $Z$, or only over a sub-group of star clusters in the same bin of mass. The merger efficiency is a good
proxy for the merger rate, because it does not depend on assumptions about the star
formation rate and metallicity evolution of the Universe. Here we explore the dependence of the merger efficiency on both $m_{\rm SC}$  and $Z$.

Figure~\ref{fig:meffsc} displays the merger efficiency versus the cluster mass for
low-mass YSCs (this work) and high-mass systems studied in \cite{dicarlo2020}. In general, the merger efficiency of the loose YSCs of set~B is smaller than the merger efficiency of the dense YSCs (set~A) with the same YSC mass. Thus, in the first approximation, a higher initial stellar density implies a higher merger efficiency.

In the loose YSCs of set~B of \cite{dicarlo2020}, the merger efficiency increases with the cluster mass by about one order of magnitude between $m_{\rm SC}=10^3$ M$_\odot$ and $3\times{}10^4$ M$_\odot$. We see a similar trend for all the considered metallicities, within the Poissonian uncertainties  (it may become weaker at $Z=0.02$, but we have low statistics).
 
 The behaviour of $\eta{}$ in the densest YSCs (those presented in this paper plus set~A of \citealt{dicarlo2020}) is more varied, depending on the metallicity. 
 At $Z=0.0002$, $\eta$ tends to increase with the cluster mass, even if the increase is less steep than for set~B. 
 At Z = $0.002$, 
 $\eta{}$ still tends to increase with cluster mass, but the trend is
 flatter than that for $Z = 0.0002$. Finally,  at solar metallicity ($Z = 0.02$), $\eta$ has a flat trend for low-mass and high-mass dense clusters. 

Furthermore, at low metallicity ($Z = 0.0002$, $0.002$), the merger efficiency of the dense clusters with $m_{\rm{SC}}>10^{4}$ \Ms{} is slightly lower than that of smaller dense YSCs ($m_{\rm{SC}}\in{}[3000,\,{}9500$] \Ms{}). This decrease is a consequence of the higher density, which leads to the break up of more binary stars. In fact, looking at the population of merging BBHs, we find that the percentage of original binaries decreases at higher star cluster mass: original BBH mergers are 85\% (75\%), 73\% (42\%) 
and 53\% (42\%) of all  BBH mergers in the second, third and fourth mass-bin, 
at Z=$0.0002$ (Z=$0.002$).
On the one hand, in most massive YSCs the number of dynamical encounters and thus the formation of
new binaries (exchanged BBHs) is significantly enhanced. On the other hand, in such massive
 systems, encounters may break  original binaries, removing a fraction 
 of potential merging BBHs from the systems.

This comparison underlines that star cluster density and star cluster mass are two crucial ingredients for the merger efficiency of BBHs in YSCs.
As discussed in \citet{dicarlo2020}, at low metallicities, where BH masses are higher, dynamical encounters
and hardening are more effective in the dense clusters of set~A than
in the loose clusters of set~B.

\subsection{Merger Efficiency and Local Merger Rate Density}

Table~\ref{tab:merg_eff} shows the merger efficiency
for BBHs from low-mass YSCs ($\eta_{\rm SC}$) and from isolated binaries ($\eta_{\rm
IB}$) at different metallicity.

\begin{table}
\centering
\begin{tabular}{@{}ccc@{}}
\toprule
$Z$    &   $\eta_{\rm SC}$   & $\eta_{\rm IB}$   \\
       &   $[{\rm M}_\odot^{-1}]$   & $[{\rm M}_\odot^{-1}]$    \\ \midrule
0.0002 &   $1.4\times{}10^{-5}$   & $1.3\times{}10^{-4}$      \\
0.002  &   $8.3\times{}10^{-6}$   & $2.9\times{}10^{-5}$      \\
0.02   &   $1.3\times{}10^{-6}$   & $3.2\times{}10^{-9}$       \\ \bottomrule
\end{tabular}
 \caption{Merger efficiency of BBHs for low-mass YSCs and isolated binaries.
 Column 1: metallicity $Z$; column 2: BBH merger efficiency for YSCs $\eta_{\rm SC}$; column 3: BBH merger efficiency for isolated binaries $\eta_{\rm IB}$.}
\label{tab:merg_eff}
\end{table}
%

In metal-poor systems ($Z=0.0002$, $0.002$),
the BBH merger efficiency for small YSCs ($\eta_{\rm SC}$) is about a factor of $\approx 9$ ($\approx 4$) lower
than the BBH merger efficiency for isolated binaries ($\eta_{\rm IB}$). 
This is an effect of dynamics, that may break up low-mass BBHs
\citep{zevin2017,dicarlo2019a} resulting in a lower merger efficiency with respect to isolated binaries.

In contrast, at solar
metallicity ($Z=0.02$) $\eta_{\rm SC}$ is
higher by two orders of magnitude than the $\eta_{\rm IB}$.
In fact, isolated BBH mergers are extremely rare at solar metallicity, because of the larger efficiency of stellar winds. In fact, the majority of massive stars become Wolf-–Rayet stars before they can start a Roche lobe event and do not undergo a common envelope phase: most of the isolated BBHs 
 that form at solar metallicity are too wide to merge within a Hubble time. 
 In contrast, dynamics triggers BBH mergers even in metal-rich YSCs: dynamical encounters lead to the formation of exchanged BBHs and, additionally, gravitational interactions harden existing massive binaries, even in metal-rich systems.

 From the merger efficiency, we can estimate the merger rate density in the
local Universe as described in \cite{santoliquido2020,santoliquido21}:
\begin{eqnarray}
\label{eq:rate}
   \mathcal{R}_{\text{BBH}} = \frac{1}{t_{\rm lb}(z_{\text{loc}})}\int_{z_{\rm max}}^{z_{\text{loc}}}\psi(z')\,{}\frac{{\rm d}t_{\rm lb}}{{\rm d}z'}\,{}{\rm d}z' \times{}\nonumber{}
   \\
   \int_{Z_{\rm min}(z')}^{Z_{\rm max}(z')}\eta{}(Z)\,{}\mathcal{F}(z',z_{\text{loc}}, Z)\,{}{\rm d}Z,
\end{eqnarray}

where $t_{\rm lb}(z_{\text{loc}})$ is the look-back time evaluated in the local
Universe ($z_{\text{loc}}\leq -2.1$), $Z_{\rm min}(z')$ and $Z_{\rm
max}(z')$ are the minimum and maximum metallicity of stars formed at redshift
$z'$, $\psi(z')$ is the cosmic SFR density at
redshift $z'$ (from \citealt{madau2017}) and $\mathcal{F}(z',z_{\text{loc}}, Z)$ is the 
fraction of BBHs that form at redshift $z'$ from stars with metallicity $Z$ and merge at redshift
$z_{\text{loc}}$ normalized to all BBHs that form from stars with metallicity
$Z$. The lookback time $t_{\rm lb}$ is estimated taking the cosmological
parameters from
\cite{ade2016}. We integrate equation~\ref{eq:rate} up to redshift $z_{\rm
max}=15$.

From equation~\ref{eq:rate} and assuming the metallicity evolution proposed by \cite{madau2017}, we obtain a local merger rate density
$\mathcal{R}_{\rm BBHs}\, =\, 88^{+34}_{-26}$ $f_{\rm YSC}$ Gpc$^{-3}$~yr$^{-1}$, 
where $f_{\rm YSC}\in (0,1]$ is the fraction of the cosmic star formation rate that occurs in YSCs 
like the ones we simulated. The uncertainty refers to the 50\% credible interval and accounts for both star formation rate and metallicity evolution. Calculating the rate for isolated BBHs with the same methodology, we found $\mathcal{R}_{\rm BBHs}\sim{}50$ $f_{\rm IB}$ Gpc$^{-3}$~yr$^{-1}$ \citep{santoliquido2020}, where $f_{\rm IB}\in [0,1]$ is the fraction of the cosmic star formation rate happening in the field.
The BBH merger rate density of our low-mass YSCs is similar to the
  BBH merger rate density of higher mass YSCs,  estimated by \cite{dicarlo2020}  ($\mathcal{R}_{\rm BBHs}\sim{}55$ Gpc$^{-3}$~yr$^{-1}$), \cite{banerjee2021} ($\mathcal{R}_{\rm BBHs}\sim{}0.5-38$ Gpc$^{-3}$ yr$^{-1}$) and \cite{kumamoto2020} ($\sim{}70$ Gpc$^{-3}$ yr$^{-1}$).  It seems unlikely that all the BBH mergers observed by the LVC come from a single formation channel \citep[see, e.g.,][]{abbottO3apopandrate,zevin2021,callister2021,wong2021,bouffanais2021}. 
The LVC recently estimated a BBH merger rate of $\mathcal{R}_{\rm BBHs}=52^{+52}_{-26}$  Gpc$^{-3}$~yr$^{-1}$ and $\mathcal{R}_{\rm BBHs}=23.9^{+14.3}_{-8.6}$  Gpc$^{-3}$~yr$^{-1}$, depending on whether we include or not GW190814 in the BBH population, respectively \citep{abbottO3apopandrate}. When compared to the BBH merger rate density inferred by the LVC, our results suggest that YSCs can give a substantial contribution to the population of BBH mergers in the local Universe.

\section{Summary and Conclusions}

We have studied the formation of BBHs in $\sim{}10^5$ low-mass young star clusters
($300$ \Ms $\,<m_{\rm{SC}}<\,10^3$ \Ms)  with different
metallicities $Z = 0.02$, $0.002$ and $0.0002$, by means of direct $N$-body
simulations. We used the code \texttt{NBODY6++GPU}
\citep{wang2015} interfaced with the population-synthesis code \texttt{MOBSE}
\citep{giacobbo2018}.
We compare our results with those of high-mass YSCs ($10^3$ \Ms $\,<m_{\rm{SC}}<\,3\times{}10^4$ \Ms) presented in 
\cite{dicarlo2019a} and \cite{dicarlo2020} and with a sample of isolated binaries evolved with the code \texttt{MOBSE} by \citet{giacobbo2018b}.

The lighter YSCs in our sample ($300<\,m_{\rm{SC}}\,<700$ \Ms) tend to over-fill their tidal radius over $\approx 100$ Myr and expand more rapidly than the heavier YSCs
($700<\,m_{\rm{SC}}\,<1000$ \Ms{}) which, in contrast, remain
tidally under-filling for the entire simulation. 
The relaxation time scale of low-mass YSCs is extremely short, $t_{\mathrm{rlx}}\approx $ few tens of Myr. Our clusters are created as already core-collapsed and thus experience a rapid initial expansion of the core radius and an initial drop of the
central density (Figure \ref{fig:9plot}). 

About $75-85$\% of the BHs formed in our simulations are ejected from their parent cluster because of dynamical recoil. For comparison, only $\sim{}35-65$\% of the stars are ejected by the end of the simulation. This difference springs from Spitzer's instability: the most massive stars (BH progenitors) segregate to the core of the star cluster over a very short timescale ($<1$ Myr) in our YSCs. In the core, they undergo dynamical encounters and eject each other, while most of the low-mass stars remain in the loose outskirts of the star cluster.

Focusing on the population of BBHs, we have shown that the typical mass of a BBH does not
depend on the cluster mass (Figure \ref{fig:mtot_bbh_mysc}). We distinguish between original BBHs, which are the result of the evolution of original binary stars, and exchanged BBHs, which form from dynamical exchanges. Exchanged BBHs have median values of the total mass ($m_{\rm BBH}=m_1+m_2$) of the order of $\sim 80$, $\sim 60$ and $\sim 20$ \Ms{} at
$Z=0.0002$, $0.002$ and $0.02$, respectively. Original BBHs are, on average, less massive ($\sim 60$, $\sim 30$ and $\sim 10$ \Ms{} at $Z=0.0002$, $0.002$ and $0.02$, respectively). The masses of original BBHs are more similar to isolated BBHs
\citep{giacobbo2018} while exchanged BBHs host primary binary components with 
mass up to $\approx 190$ \Ms{} (Figure \ref{fig:m1_m2_tot}).

 We find that the percentage of exchanged BBHs strongly depends on the star cluster mass: it increases from $\sim 40-60$ \% in the lowest-mass YSCs ($300$ M$_\odot$) up to more than
$\approx 75-95$ \% in the most massive YSCs ($3\times{}10^4$ M$_\odot$, from set~A of \citealt{dicarlo2020}). The large number of dynamical scatterings in the most massive clusters enhances the probability to form exchanged BBHs and, at the same time, facilitates the ionization of original binaries.  

We then focused on the sub-sample of BBHs that reach coalescence in less than a Hubble time.
BBH mergers in low-mass YSCs tend to be relatively light, $m_{\rm{BBH}}\,<\,70$ \Ms.
About $\approx 90\%$ of merging BBHs are original binaries 
and most of them (60\%) are produced in metal poor environments. 
Exchanged BBH mergers in low-mass YSCs have masses similar to those of original BBHs. This is very different from what \cite{dicarlo2020} found for more massive clusters ($>10^3$ M$_\odot$), where exchanged BBH mergers are significantly more massive than original BBH mergers.  
In low-mass YSCs (Figure \ref{fig:tex}),  most of the exchanged BBHs pair up before the formation of the two BHs: the dynamical exchange involves the progenitor stars. If the binary semi-major axis at time of the exchange is sufficiently tight, the binary experiences
mass transfer and/or common envelope before the formation of the second BH, causing
 mass loss from the progenitor stars that leads to the
formation of light BHs. 
Moreover, very massive BH progenitors tend to form binaries with large semi-major axis ($a \approx 10^{3}$ AU) that are difficult to shrink trough binary evolution processes or dynamical interactions, which become rather rare as the clusters evolve (Figure \ref{fig:9plot}).  

Mass ratios $q\,<\,0.4$ are generally more common in dynamical merging BBHs
(both original and exchanged BBHs) than in isolated binaries, because dynamics may
pair up BHs with different masses. In low-mass YSCs the majority of coalescing BBHs merge after the binaries have been ejected by the parent clusters before $\approx 2$ Gyr.

The BBH survival
efficiency $\eta{}_{\rm s}$, i.e. the number of BBHs that form and survive to the end of the simulation, divided by the total initial star cluster mass, is almost independent of the star cluster mass and central density, especially in metal poor-environments. At solar metallicity, there is a different trend when considering massive dense or loose YSCs (respectively set A and set B of \citealt{dicarlo2020}).

In loose YSCs, the merger efficiency $\eta$ (i.e. the number of BBH mergers divided by the total initial star cluster mass) increases with the cluster mass by up to one order of magnitude. In dense clusters $\eta$ and the cluster mass show a very mild correlation in metal poor environments and no correlation at solar metallicity. 

At low metallicity ($Z=0.0002$), the merger efficiency of YSCs is about one order of magnitude lower than that of isolated binaries, because dynamics ejects BHs from the cluster and might break binary systems.
In contrast,  at $Z=0.02$, the merger efficiency of YSCs is $\sim{}400$ times higher than that of isolated binaries. In fact, unperturbed binary stars at high metallicity are very difficult to merge, because the progenitor stars become Wolf-Rayet systems (with small stellar radii) and cannot undergo common envelope episodes: their semi-major axis remains too large. In YSCs, dynamics triggers BBH mergers even in metal-rich YSCs. 
From the merger efficiency \citep{santoliquido2020}, we estimated a local BBH merger rate density of low-mass YSCs  $\mathcal{R}_{\rm BBHs}\, =\,88^{+34}_{-26}$ $f_{\rm YSC}$ Gpc$^{-3}$~yr$^{-1}$, where $f_{\rm YSC}\in[0,1]$ is the fraction of the cosmic star formation rate that occurs in YSCs. This result
suggests that low-mass YSCs have a significant role in the population of BBH mergers in the local Universe.

\section*{Acknowledgements} 
MM, SR, UNDC, GI, AB, NG and FS acknowledge financial support by the
European Research Council for the ERC Consolidator grant DEMOBLACK, under
contract no. 770017. NG acknowledges financial support  by  Leverhulme  Trust  Grant  No.  RPG-2019-350 and Royal Society Grant No. RGS-R2-202004.  We are grateful to Long Wang for
 his helpful support on {\sc nbody6++gpu}.
Part of the $N$-body simulations discussed in this paper
were performed at the supercomputer DEMOBLACK at the Physics and Astronomy
department \vir{G. Galilei} of the University of Padova, equipped with 192 dual
cores, 8 V100 NVIDIA GPUs. We acknowledge the CINECA-INFN agreement, for the availability of high performance computing resources and support.

\section*{Data availability}
The data underlying this article will be shared on reasonable request to the corresponding authors.

\bibliographystyle{mnras}
\bibliography{rasetal}

\end{document}